\documentclass[twocolumn,superscriptaddress,prb,10pt]{revtex4-1}
\usepackage{verbatim}
\usepackage{amsmath,amssymb}
\usepackage{graphicx}
\usepackage{color}
\usepackage[colorlinks,bookmarks=false,citecolor=blue,linkcolor=red,urlcolor=blue]{hyperref}
\usepackage{times}

\def\tr{\textrm{Tr}}

\begin{document}

\title{The classical mutual information in mean-field spin glass models} 

\author{Vincenzo Alba}
\affiliation{International School for Advanced Studies (SISSA),
Via Bonomea 265, 34136, Trieste, Italy, 
INFN, Sezione di Trieste}
\author{Stephen Inglis}
\affiliation{Department of Physics and Arnold Sommerfeld
Center for Theoretical Physics, Ludwig-Maximilians-Universit\"at
M\"unchen, D-80333 M\"unchen, Germany}
\author{Lode Pollet}
\affiliation{Department of Physics and Arnold Sommerfeld
    Center for Theoretical Physics, Ludwig-Maximilians-Universit\"at
M\"unchen, D-80333 M\"unchen, Germany}

\date{\today}

\begin{abstract} 

We investigate the \emph{classical} R\'enyi entropy $S_n$ and the associated mutual 
information ${\mathcal I}_n$ in the Sherrington-Kirkpatrick (S-K) model, which is the 
paradigm model of mean-field spin glasses. 
Using classical Monte Carlo simulations and analytical tools we investigate the 
S-K model on the $n$-sheets booklet.
This is obtained by gluing together $n$ independent copies of the model, and it is the 
main ingredient to construct the R\'enyi entanglement-related quantities. We find a 
glassy phase at low temperature, whereas at high temperature the model exhibits 
paramagnetic behavior, consistent with the regular S-K model. The temperature of the 
paramagnetic-glassy transition depends 
non-trivially on the geometry of the booklet. At high-temperatures we provide the exact 
solution of the model by exploiting the replica symmetry. This is the permutation symmetry 
among the fictitious replicas that are used to perform disorder averages (via the replica 
trick). In the glassy phase the replica symmetry has to be broken. Using a generalization 
of the Parisi solution, we provide analytical results for $S_n$ and ${\mathcal I}_n$, 
and for standard thermodynamic quantities. Both $S_n$ and ${\mathcal I}_n$ exhibit a 
volume law in the whole phase diagram. We characterize the behavior of the 
corresponding densities $S_n/N,{\mathcal I}_n/N$, in the thermodynamic limit. 
Interestingly, at the critical point the mutual information does not exhibit any 
crossing for different system sizes, in contrast with local spin models. 

\end{abstract}


\maketitle

\section{Introduction}

Besides being ubiquitous in nature, disorder leads to several intriguing 
physical phenomena. 
Arguably, \emph{spin glasses} represent one of the most prototypical examples
of interesting behavior induced by disorder.
While at any finite temperature disorder can prevent the 
usual magnetic ordering, at low-enough temperatures these systems display a 
new type of ``order''. In the past decades an intense theoretical 
effort has been devoted to characterizing this spin glass order, the nature of 
the paramagnetic-glassy transition, and that of the associated order 
parameter~\cite{binder-1986,parisi-book,young-1998,nishimori-book,castellani-2005}. 

All these issues can be thoroughly addressed in the Sherrington-Kirkpatrick 
(S-K) model~\cite{sherrington-1978,sherrington-1978-prl}, which is exactly 
solvable. The S-K model is a \emph{classical} Ising model on the fully-connected 
graph of $N$ sites, with quenched random interactions. Its hamiltonian reads  
\begin{equation}
{\mathcal H}=-\sum\limits_{1\le i<j\le N}J_{ij}S_i S_j-
h\sum\limits_{1\le i\le N}S_i.
\label{SK-intro}
\end{equation}
Here $S_i=\pm 1$ are classical Ising spins, $h$ is an external magnetic field, and 
$J_{ij}$ are uncorrelated (from site to site) random variables. The S-K model hosts 
a low-temperature glassy phase, which is separated from a high-temperature paramagnetic 
one by a second order phase transition. Despite its mean-field nature, the solution 
of the S-K model has been a mathematical challenge. Although it was proposed as an 
ansatz by Parisi~\cite{parisi-1980} more than thirty years ago, its rigorous proof 
was obtained only recently~\cite{talagrand-2006}. Moreover, the solution exhibits several 
intricate features, such as lack of self-averaging~\cite{pastur-1991}, ultrametricity~\cite{
mezard-1984,rammal-1986}, and replica symmetry breaking~\cite{parisi-book,castellani-2005}. 
The last refers to the breaking of the permutation symmetry among the fictitious replicas 
of the model, which are introduced to perform disorder averages (via the so-called 
\emph{replica trick}~\cite{cardy-book}. Finally, although the 
applicability of the S-K model to describe realistic spin glasses is still highly 
debated~\cite{yucesoy-2012,billoire-2012,yucesoy-2013}, there are recent proposals on 
how to realize it in cold-atomic gases~\cite{morrison-2008,rotondo-2015}, or in laser 
systems~\cite{ghofraniha-2015}. 

In the last decade entanglement-related quantities emerged as valuable tools to 
understand the physics of complex systems~\cite{amico-2008,eisert-2009,calabrese-2009,
cc-rev}, both classical and quantum. For instance, at a conformally invariant critical 
point entanglement measures contain universal information about the underlying 
conformal field theory (CFT), such as the central charge~\cite{holzhey-1994,vidal-2003,
calabrese-2004,calabrese-2012}. For classical spin models a lot of attention has been 
focused on the \emph{classical} R\'enyi entropy~\cite{jaconis-2013,stephan-2014}. Given 
a bipartition of the system into two complementary subregions $A$ and $B$, the classical 
R\'enyi entropy $S_n(A)$ (with $n\in\mathbb{N}$) is defined as 
\begin{equation}
S_n(A)\equiv \frac{1}{1-n}\log\Big(\sum\limits_{i_A\in{\mathcal C}_A} p^n_{i_A}
\Big)
\label{renyi-intro}
\end{equation}
Here ${\mathcal C}_A$ denotes the set of all the possible spin configurations in part 
$A$, whereas $p_{i_A}$ is the probability of the configuration $i_A$. 
Alternatively, $S_n(A)$ can be obtained from the partition function of the model on 
an \emph{ad hoc} defined ``booklet'' geometry (see section~\ref{booklet} for its 
definition). This consists of $n$ independent and identical copies (the 
booklet ``sheets'') of the model. These \emph{physical} copies are different 
from the \emph{fictitious} replicas used to perform the disorder average. Each sheet is 
divided into two parts $A$ and $B$, containing $N_A$ and $N_B$ spins, respectively. 
The spins in part $A$ of different sheets are constrained to be equal. It is 
convenient to introduce the booklet aspect ratio $0\le\omega\le1$ as  
\begin{equation}
\label{a-ratio}
\omega\equiv \frac{N_A}{N}.
\end{equation}
%
For local spin models the bipartition correspond to a spatial separation 
between the spins. However, the definition~\eqref{renyi-intro} can be used  
in models with no notion of space, where it quantifies the correlation between two 
groups of spins rathen than two spatial regions.  
Notice that Eq.~\eqref{renyi-intro} can also be used for quantum systems by replacing 
the sum over the proababilities of each state in region $A$ with the trace over the reduced 
density matrix for region $A$. For $n=1$ 
Eq.~\eqref{renyi-intro} defines the subsystem Shannon entropy~\cite{alcaraz-2013,
stephan-2014-a}. From $S_n(A)$, one defines the classical mutual information 
${\mathcal I}_n(A,B)$ as 
\begin{equation}
\label{MI}
{\mathcal I}_n(A,B)\equiv S_n(A)+S_n(B)-S_n(A\cup B). 
\end{equation}
For local models ${\mathcal I}_n$ obeys the area law~\cite{wolf-2008} ${\mathcal I}_n(A)
\propto\ell$, with $\ell$ the length of the boundary between $A$ and $B$. Remarkably, 
for different $\ell$, the ratio ${\mathcal I}_n/\ell$ has been shown to exhibit a crossing at a second 
order phase transition~\cite{jaconis-2013}, implying that it can be used as a diagnostic 
tool for critical behaviors. For conformally invariant critical models more universal 
information can be extracted from the area-law corrections of ${\mathcal I}_n$~\cite{stephan-2014}. 

Although recently the study of the interplay between disorder and entanglement became a fruitful 
research area~\cite{refael-2009}, the behavior of entanglement-related quantities 
in glassy phases, and at glassy critical points, has not been explored yet (see, however,  
Ref.~\onlinecite{castelnovo-2010} for some interesting results). Here we investigate 
both the classical R\'enyi entropy $S_n(A)$ and the mutual information ${\mathcal I}_n$ 
in the S-K model, using classical Monte Carlo simulations and analytical tools. We often 
restrict ourselves to the case with $n=2$, as this is where numerical simulations are 
most efficient. As usual in disordered system, we focus on disorder-averaged quantities, 
considering $[S_n]$ and $[{\mathcal I}_n]$, with the brackets $[\cdot]$ denoting the 
average over different realizations of $J_{ij}$ (cf. Eq.~\eqref{SK-intro}).

\begin{figure}[t]
\includegraphics*[width=0.93\linewidth]{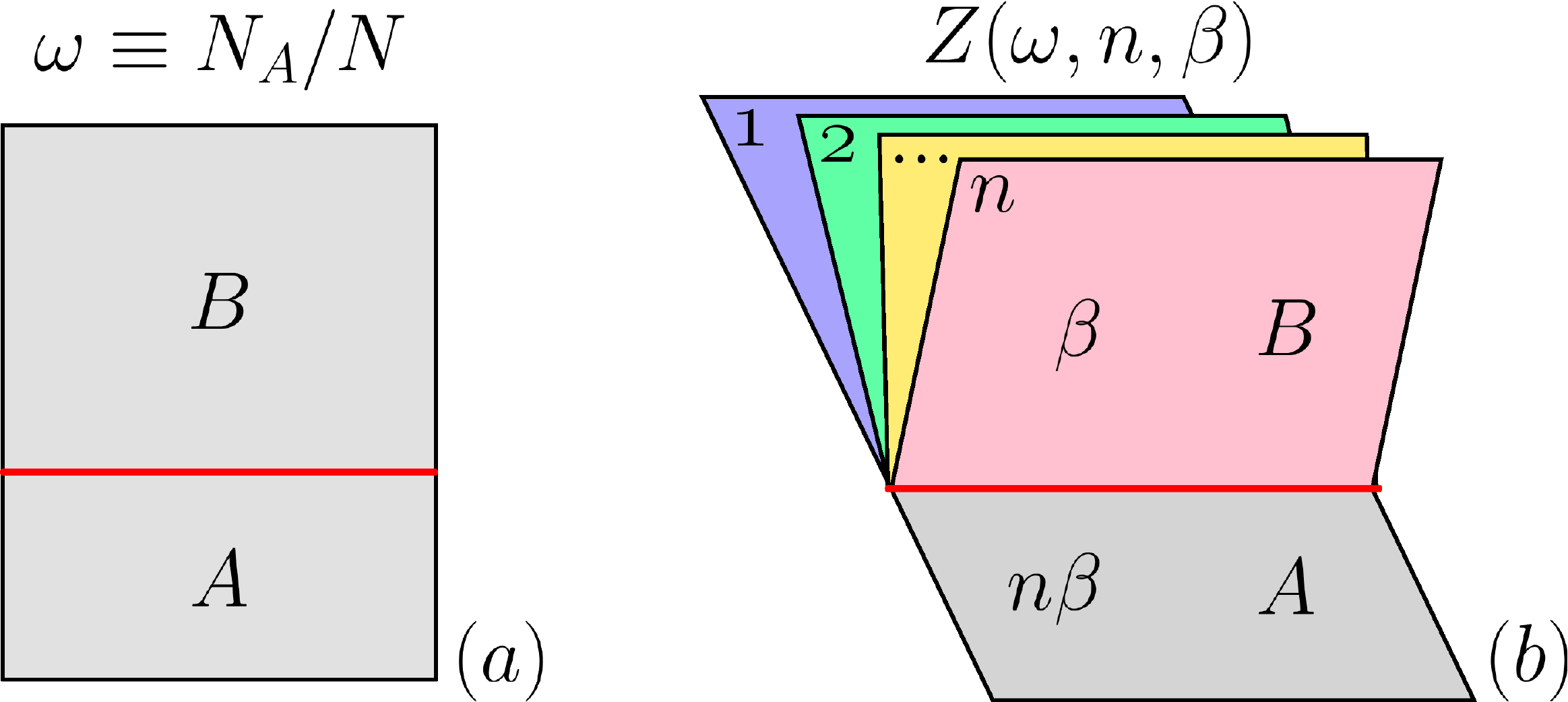}
\caption{ The booklet geometry considered in this work. (a) The single sheet 
 (``page'') of the booklet: The $N$ spins living on the sheet are divided into 
 two groups $A$ and $B$, containing $N_A$ and $N_B$ spins, respectively. Here 
 $\omega\equiv N_A/N$ is the booklet ratio. (b) The $n$ sheets are glued together 
 to form the booklet. The spins in part $B$ of the booklet pages are at inverse 
 temperature $\beta$. The spins in part $A$ are identified  (see Eq.~\eqref{book-constraint}). 
 As a consequence, the effective temperature in part $A$ is $n\beta$. 
 $Z(\omega,n,\beta)$ denotes the partition function of the S-K model on 
 the booklet. 
}
\label{cartoon}
\end{figure}

The Article is organized as follows. In section~\ref{booklet} we introduce the classical 
R\'enyi entropy and the mutual information, reviewing their representation in terms of 
the booklet partition functions. In section~\ref{the-model} we present the structure of 
the solution of the S-K model on the booklet. Specifically, we discuss the replica 
trick, which is used to perform disorder averages, and the saddle point approximation in 
the thermodynamic limit. Section~\ref{solution} is concerned with the RS approximation. 
In subsection~\ref{para-section} we focus on the high-temperature region, where this 
approximation becomes exact. In subsection~\ref{rs-section} we discuss the structure of 
the RS ansatz in the low-temperature region. Section~\ref{rsb-1-section} is devoted to 
the $1$-RSB approximation. In section~\ref{mc-results} we check the validity of both the 
RS and the $1$-RSB results comparing with Monte Carlo simulations for the internal energy. 
Section~\ref{Renyi-section} and section~\ref{I2-section} discuss the classical R\'enyi 
entropy and the mutual information, respectively. We conclude in section~\ref{conclusions}.

\section{The booklet construction \& and the classical R\'enyi entropy}
\label{booklet}

Given a generic \emph{classical} spin model at inverse temperature $\beta$, 
The probability $p_i$ of a given spin configuration $i\in{\mathcal C}$, with ${\mathcal 
C}$ being the set of all possible configurations, is given by the Boltzmann weight 
$p_i=e^{-\beta E(i)}/Z$, with $E(i)$ the associated energy. In presence of a 
bipartition (see Fig.~\ref{cartoon} (a)) the probability of a spin configuration 
$i_A\in{\mathcal C}_A$ is given as 
$p_{i_A}=\sum_{i_B}e^{-\beta E(i_A,i_B)}/Z$, where the sum is over all possible 
spin configurations in part $B$. Note that this is valid for generic interactions, 
i.e., both local and non-local ones. Clearly, one has $p^n_{i_A}\equiv\frac{1}{Z^n}
\sum_{(i^1_B,\dots,i^n_{B})}e^{-\beta\sum_k E(i_A,i^k_{B})}$. 
Thus, the classical R\'enyi entropy $S_n(A)$ (Eq.~\eqref{renyi-intro}) can be 
calculated as~\cite{jaconis-2013,stephan-2014} 
\begin{equation}
\label{renyi}
S_n(A)\equiv \frac{1}{1-n}\log\left(\frac{Z(A,n,\beta)}{Z^n(\beta)}\right).
\end{equation}
Here $Z(A,n,\beta)\equiv\sum_{i_A,(i_B^1,\dots,i_B^n)}e^{-\beta\sum_k E(i_A,i^k_{B})}$ 
can be interpreted as  the partition function of the model on the $n$-sheet booklet (with $k$ 
labelling its different sheets), whereas $Z(\beta)$ is the partition functions on the plane at 
inverse temperatures $\beta$. The booklet geometry is illustrated in Fig.~\ref{cartoon}, 
and consists of $n$ identical copies (``sheets'') of the system. Each sheet is divided into two 
parts $A$ and $B$ (cf. Fig.~\ref{cartoon} (a)). The spins in part $A$ of the different 
sheets are identified (cf. Fig.~\ref{cartoon} (b)). While spins in parts $B$ of the booklet 
are at inverse temperature $\beta$, the ones in $A$ are at the effective temperature $n\beta$. 
Notice that a similar geometric construction~\cite{guerra-2002} plays an important role in the 
mathematical proof of the Parisi ansatz. Moreover, the study of the partition function of 
critical models on the booklet has attracted considerable attention recently~\cite{fradkin-2006,
fradkin-2009,hsu-2009,stephan-2009,hsu-2010,oshikawa-2010,stephan-2010,zaletel-2011}.
Clearly, one has $S_n(A)\equiv1/(n-1)(\log(F(A,n,\beta))-n\log(F(\beta)))$, where $F(A,n,\beta)
\equiv\log(Z(A,n\beta))$ and $F(\beta)\equiv\log(Z(\beta))$ are related (apart from a factor 
$-1/\beta$) to the free energy of the model on the booklet and on the plane, respectively.

The mutual information ${\mathcal I}_n(A,B,\beta)$ is obtained in terms of the booklet 
partition functions using Eq.~\eqref{renyi} and Eq.~\eqref{MI}, where $S_n(B)$ is 
obtained from~\eqref{renyi} by exchanging $A$ and $B$. Notice also that in~\eqref{renyi}  
$Z(A\cup B,n,\beta)\equiv Z(n\beta)$ is the partition function of the model on the booklet 
with all the $n$ sheets identified, equivalently on a single sheet but at temperature 
$n\beta$. 
Notice that the disorder-averaged 
mutual information $[{\mathcal I}_n]$ and the R\'enyi entropy $[S_n]$ are directly related to 
the so-called quenched-averaged free energy $[F(A,n,\beta)]$, which is the main quantity of 
interest in disordered systems~\cite{cardy-book}. 
For clean (i.e., without disorder) \emph{local} spin models ${\mathcal I}_n$ 
obeys the boundary law 
\begin{equation}
{\mathcal I}_n(A,B,\beta)=\alpha_n\ell+{\mathcal G}_n+\gamma_n,
\label{GMI}
\end{equation}
with $\ell$ the length of the boundary between $A$ and $B$, and 
$\alpha_n,\gamma_n$ two non-universal constants. Here ${\mathcal G}_n$ 
is the so-called geometric mutual information~\cite{stephan-2014}. 
Interestingly, for critical systems ${\mathcal G}_n$ depends only on the 
geometry of $A$ and $B$, and it is universal. For conformally invariant models 
${\mathcal G}_n$ can be calculated using standard methods of conformal field 
theory (CFT), and it allows to numerically extract universal information 
about the CFT, such as the central charge~\cite{stephan-2014}.

\section{The Sherrington-Kirkpatrick (S-K) model on the booklet}
\label{the-model}

Here we introduce the Sherrington-Kirkpatrick (S-K) model on the booklet. 
In subsection~\ref{the-model-def}  we define the 
model and its partition function. In subsection~\ref{replica-sec} 
we discuss the replicated booklet construction that is used to calculate the 
disorder-averaged free energy $[F(\omega,n,\beta)]$. 
In section~\ref{saddle-point} we consider the thermodynamic limit, 
using the saddle point approximation. We also introduce the overlap tensor,  
which contains all the information about the thermodynamic behavior of the 
model. The analytical formula for the replicated partition function (Eq.~\eqref{Z-ac}), 
and the saddle point equations (Eqs.~\eqref{saddle-final}\eqref{saddle-final-a}) 
for the overlap tensor are the main results of this section. 

\subsection{The model and its partition function}
\label{the-model-def}

The Sherrington-Kirkpatrick (S-K) model~\cite{sherrington-1978-prl,
sherrington-1978} on the $n$-sheets booklet (cf. Fig.~\ref{cartoon}) is 
defined by the Hamiltonian
\begin{equation}
{\mathcal H}=-\sum\limits_{r=1}^n\left\{\sum\limits_{i<j}J_{ij}S^{(r)}_i 
S^{(r)}_j-h\sum\limits_{i=1}^NS^{(r)}_i\right\}.
\label{SK-ham}
\end{equation}
Here $S_i^{(r)}=\pm 1$ are classical Ising spins, $r\in[1,n]$ labels the 
different sheets (``pages'') of the booklet, $i\in[1,N]$ denotes the sites on 
each sheet, $J_{ij}$ is the interaction strength, and $h$ is an external 
magnetic field. The total number of spins in the booklet is $nN$. 
The first sum inside the brackets in Eq.~\eqref{SK-ham} is over all the 
$N(N-1)/2$ pairs of spins in each sheet. Spins on different sheets do not interact. 
In each ``sheet'' all the sites are divided into two groups $A\equiv\{1,\dots, N_A\}$ 
and $B\equiv\{N_A+1,\dots,N\}$, containing $N_A$ and $N_B\equiv N-N_A$ sites, 
respectively. The spins living in part $A$ and different sheets are identified, i.e. 
one has  
\begin{equation}
S_i^{(r)}=S_{i}^{(r')}\quad\forall\, i\in A,r,r'. 
\label{book-constraint}
\end{equation}
Since in each sheet all spins interact with each other, there is no notion of distance 
between different spins. Thus, physical observables should depend on the booklet geometry 
only through the ratio $\omega$ (cf. Eq.~\eqref{a-ratio}). 

In Eq.~\eqref{SK-ham} the couplings $J_{ij}\in\mathbb{R}$ are uncorrelated (from site to site) 
quenched random variables. $J_{ij}$ are the same in all the sheets of the booklet. Specifically, 
here $J_{ij}$ are drawn from the gaussian distribution   
\begin{equation}
P(J_{ij})=
\left(\frac{N}{2\pi}\right)^{1/2}
\exp\Big\{-\frac{N}{2J^2}
\Big(J_{ij}-\frac{J_0}{N}\Big)^2\Big\}.
\label{quenched-distr}
\end{equation}
The mean and the variance of $P(\{J_{ij}\})$ are given as $[J_{ij}]=J_0/N$ 
and $[(J_{ij}-[J_{ij}])^2]=J^2/N$, respectively. Here we set $J=1$. The square brackets $[\cdot]$ denote 
the average over different realizations of $J_{ij}$. Here we restrict ourselves to 
$J_0=0$ and $J=1$. The factors $N$ in Eq.~\eqref{quenched-distr} ensure a well-defined 
thermodynamic limit. 

The partition function $Z(\omega,n,\beta,\{J\})$ of the S-K model on the booklet 
at inverse temperature $\beta\equiv 1/T$, and for fixed disorder realization 
$\{J_{ij}\}$, reads 
\begin{multline}
Z(\omega,n,\beta,\{J\})\equiv\textrm{Tr}'\exp(-\beta{\mathcal H})=\\
\textrm{Tr}'\exp\Big\{\beta\sum\limits_{r}\Big(
\sum\limits_{i<j}J_{ij}S^{(r)}_i 
S^{(r)}_j-h\sum\limits_{i=1}^NS^{(r)}_i\Big)\Big\},
\label{book-pf}
\end{multline}
where $\textrm{Tr}'\equiv \sum_{\{S_i\}}$ denotes the sum over all possible 
spin configurations. The prime in $\textrm{Tr}'$ stresses that only spin configurations 
satisfying the constraint in Eq.~\eqref{book-constraint} are considered. In the two limits 
$\omega=0$ and $\omega=1$ one recovers the standard S-K model. In particular, for $\omega=0$, 
i.e., $n$ disconnected sheets, one has $Z(0,n,\beta,\{J\})=Z(\beta,\{J\})^n$, with 
$Z(\beta,\{J\})$ the partition function of the S-K model on the plane (i.e., the original 
S-K model). On the other hand, 
for $\omega=1$ it is $Z(1,n,\beta,\{J\})=Z(n\beta,\{J\})$, i.e., the partition function 
of the S-K model at inverse temperature $n\beta$. The quenched averaged free energy 
$[F(\omega,n,\beta)]$, is defined as 
\begin{equation}
\label{free-energy}
[F(\omega,n,\beta)]\equiv -\frac{1}{\beta}\int {\mathcal D}\{J\}
\log Z(\omega,n,\beta,\{J\}), 
\end{equation}
where $\int{\mathcal D}\{J\}\equiv\prod\nolimits_{i<j}\int_{-\infty}^{
+\infty}dJ_{ij}P(J_{ij})$. 

At $\omega=0$ and $\omega=1$ the phase diagram of the S-K model in the 
thermodynamic limit is well established~\cite{parisi-book,nishimori-book}. 
At $\omega=0$ it exhibits a standard paramagnetic phase in the high temperature 
region, whereas at low temperatures a glassy phase is present, with replica-symmetry 
breaking. The two phases are divided by a second order phase transition at 
$\beta=\beta_c=1$. The phase diagram for $\omega=1$ is the same, apart from the 
trivial rescaling $\beta\to n\beta$. We anticipate here (see section~\ref{rs-section}) 
that a similar scenario holds for generic $\omega$. Specifically, the glassy 
replica-symmetry-broken phase at low temperatures survives for generic 
$0<\omega<1$, while at high enough temperature the model is paramagnetic. The 
critical point, which marks the transition between the two phases, is a 
nontrivial function of the booklet ratio $\omega$ (see section~\ref{tc-section}). 

\subsection{The replicated booklet and the overlap tensor}
\label{replica-sec}

The disorder-averaged free energy $[F(\omega,n,\beta)]$ (cf. 
Eq.~\eqref{free-energy}) of the model is obtained, using the standard replica 
trick~\cite{cardy-book}, as  
\begin{equation}
\label{replica-trick}
[F(\omega,n,\beta)]=\lim_{\alpha\to 0}\frac{[Z^\alpha(\omega,n,\beta)]-1}
{\alpha}. 
\end{equation}
Here $[Z^\alpha(\omega,n,\beta)]$ is the disorder-averaged partition 
function of $\alpha\in\mathbb{N}$ independent copies of the S-K model on the 
booklet. Precisely, $[Z^\alpha(\omega,n,\beta)]$ reads 
\begin{multline}
[Z^\alpha(\omega,n,\beta)]=\int{\mathcal D}\{J\}
\tr'\exp\sum\limits_{r,\gamma}
\Big\{\\
\beta\sum\limits_{i<j}J_{ij}S_i^{(r,\gamma)}S_j^{(r,\gamma)}
+\beta h\sum\limits_{i}S^{(r,\gamma)}_i\Big\},
\label{rep-Z}
\end{multline}
where the index $\gamma=1,2,\dots,\alpha$ labels the different fictitious 
replicas introduced in Eq.~\eqref{replica-trick}, whereas $r$ denote the 
physical copies, i.e., the booklet sheets, as in Eq.~\eqref{book-pf}. Again, 
spins on different sheets or different replicas do not interact with each other. 
Notice that $Z^{\alpha}(\omega,n,\beta)$ can be thought of as the partition 
function of the S-K model on a ``replicated'' booklet. 

Using Eq.~\eqref{quenched-distr}, the disorder average in Eq.~\eqref{rep-Z} 
can be performed explicitly, to obtain 
\begin{multline}
[Z^\alpha(\omega,n,\beta)]=\tr'\exp\sum\limits_{r,\gamma}\Big\{\\
\frac{1}{N}
\sum\limits_{i<j}\Big(
\frac{\beta^2}{2}\sum\limits_{\gamma',r'}S^{(r,\gamma)}_iS^{(r,\gamma)}_j
S^{(r',\gamma')}_iS^{(r',\gamma')}_j\\
+\beta J_0 S_i^{(r,\gamma)}S_j^{(r,\gamma)}\Big)
+\beta h\sum\limits_{i}
S_i^{(r,\gamma)}\Big\}.
\label{dis-aver-Z}
\end{multline}
In contrast with Eq.~\eqref{rep-Z}, both different sheets and different replicas 
are now coupled by a four-spin interaction. It is convenient to introduce the  
Hubbard-Stratonovich variables $q_{\gamma\gamma'}^{rr'}$ and $m_\gamma^r$. 
Following the spin glass literature~\cite{parisi-book}, we dub $q_{\gamma
\gamma'}^{rr'}$ the overlap tensor. In the standard S-K model (i.e., for $n=1$) 
$q_{\gamma\gamma'}^{rr'}$ becomes a $\alpha\times\alpha$ 
matrix~\cite{sherrington-1978-prl}. Eq.~\eqref{dis-aver-Z} now 
yields  
\begin{multline}
\label{hs-Z}
[Z^\alpha(\omega,n,\beta)]=\exp\Big(\frac{\beta^2Nn\alpha}{4}\Big)
\int\prod_{\substack{\gamma\le\gamma'\\r,r'}}
dq^{rr'}_{\gamma\gamma'}
\int\prod_{\gamma,r}dm_\gamma^r\\
\tr'\exp
\Big\{-N {\mathcal K}(\{q,m\})
+\sum_i{\mathcal L}_i(\{q,m\})\Big\},
\end{multline}
where we neglected subleading contributions ${\mathcal O}(1/N)$ in 
the thermodynamic limit. Here ${\mathcal K}(\{q,m\})$ is spin-independent 
and it reads 
\begin{multline}
{\mathcal K}(\{q,m\})\equiv 
\frac{\beta^2}{2J_0}\Big(\sum\limits_{\gamma<\gamma'}\sum
\limits_{r,r'} (q_{\gamma\gamma'}^{rr'})^2
\\
+\sum\limits_{\gamma}\sum\limits_{r<r'}(q_{\gamma\gamma}^{
rr'})^2\Big)
+\frac{\beta}{2J_0}\sum\limits_{\gamma r}(m_\gamma^r)^2. 
\label{Gamma}
\end{multline}
On the other hand ${\mathcal L}_i(\{q,m\})$ depends on the spin degrees of 
freedom, and it is given as   
\begin{multline}
{\mathcal L}_i(\{q,m\})
\equiv\beta^2\sum\limits_{\gamma<\gamma'}\sum\limits_{r,r'}
q_{\gamma\gamma'}^{rr'}S_i^{(r,\gamma)}
S_i^{(r',\gamma')}+\\
\beta^2\sum_\gamma\sum\limits_{r<r'}
q_{\gamma\gamma}^{rr'}S^{(r,\gamma)}_i
S^{(r',\gamma)}_i
+\beta\sum\limits_{\gamma r}(m_\gamma^r+h)
S_i^{(r,\gamma)}.
\label{mf-action}
\end{multline}
Interestingly, ${\mathcal L}_i$ describes a system of $n\alpha$ spins living 
in the replica space with the long-range interaction $q_{\gamma\gamma'}^{rr'}$, 
and a magnetic field $m_\gamma^r+h$. Notice that, while the first term in 
Eq.~\eqref{mf-action} is off-diagonal in the space of the fictitious replicas, 
the second one is diagonal. We anticipate here that the latter fully determines the 
behavior of the model in the paramagnetic phase (see section~\ref{para-section}). 

Since in Eq.~\eqref{hs-Z} spins on  different sites are decoupled, one can 
perform the trace over the spins in parts $A$ and $B$ (see Fig.~\ref{cartoon}) 
independently, to obtain 
\begin{multline}
[Z^\alpha(\omega,n,\beta)]=
\int\prod_{\substack{\gamma\le\gamma'\\r,r'}}
dq^{rr'}_{\gamma\gamma'}
\int\prod_{\gamma,r}dm_\gamma^r
\exp\Big\{N\Big(\\
\frac{\beta^2n\alpha}{4}+\omega\log\tr_Ae^{{\mathcal L}}+
(1-\omega)\log\tr_Be^{{\mathcal L}}-{\mathcal K}\Big)\Big\}.
\label{z-final}
\end{multline}
Here to lighten the notation we drop the dependence on the coordinate $i$ 
and the arguments of ${\mathcal L}_i(\{q,m\})$ and ${\mathcal K}(\{q,m\})$. 
$\textrm{Tr}_{A}$ and $\textrm{Tr}_{B}$  denote the trace over the spin degrees 
of freedom living in parts $A$ and $B$ of the booklet.
The subscript $A$ in $\tr_A$ is to stress that spins living in different  
sheets (i.e., for $r\ne r'$ in Eq.~\eqref{mf-action}) are identified (due 
to the booklet constraint in Eq.~\eqref{book-constraint}), whereas they have 
to be treated as independent variables in performing $\tr_B$.  

\subsection{The saddle point approximation}
\label{saddle-point}

In the thermodynamic limit, i.e., for $N,N_A\to\infty$, at fixed ratio 
$\omega\equiv N_A/N$, one can take the saddle point approximation in 
Eq.~\eqref{z-final}, which yields   
\begin{multline}
[Z^\alpha(\omega,n,\beta)]\approx
\exp\Big\{N\alpha\Big(\frac{\beta^2n}{4}-\frac{\mathcal K}{\alpha}\\
+\frac{\omega}{\alpha}\log\tr_A\exp({\mathcal L})+
\frac{1}{\alpha}(1-\omega)\log\tr_B\exp({\mathcal L})
\Big)\Big\}. 
\label{Z-ac}
\end{multline}
The overlap tensor $q_{\gamma\gamma'}^{rr'}$ and $m_\gamma^r$ are 
determined by solving the saddle point equations 
\begin{align}
\label{saddle}
& \frac{\partial}{\partial q_{\gamma\gamma'}^{rr'}}
\left(\omega\log\tr_A e^{{\mathcal L}}+(1-\omega)\log\tr_B 
e^{{\mathcal L}}\right)=q_{\gamma\gamma'}^{rr'}\\
\label{saddle-a}
& \frac{\partial}{\partial m_\gamma^r}\left(\omega\log\tr_A 
e^{{\mathcal L}}+(1-\omega)\log\tr_B e^{{\mathcal L}}\right)=
J_0^{-1}m_\gamma^r.
\end{align}
It is enlightening to rewrite Eqs.~\eqref{saddle}~\eqref{saddle-a} as 
\begin{align}
\label{saddle-final}
& q_{\gamma\gamma'}^{rr'}=
\omega\langle S^{(r,\gamma)} S^{(r',\gamma')}\rangle_A+
(1-\omega)\langle S^{(r,\gamma)}S^{(r',\gamma')}
\rangle_B\\
& J_0^{-1}m_\gamma^r=\omega\langle S^{(r,\gamma)}
\rangle_A+(1-\omega)\langle S^{(r,\gamma)}
\rangle_B
\label{saddle-final-a}
\end{align}
where $\langle {\mathcal O}\rangle_{A(B)}\equiv (Z_{A(B)})^{-1}
\tr_{A(B)}\{{\mathcal O}\exp({\mathcal L})\}$ with $Z_{A(B)}\equiv\tr_{A(B)}
\exp({\mathcal L})$. Notice that Eq.~\eqref{saddle-final} implies 
that $q_{\gamma\gamma}^{rr}=1$ $\forall\gamma,r$. For $\omega=0$ and $n=1$, 
one recovers the saddle point equations for the standard SK model~\cite{parisi-book,
nishimori-book}. 

In order to calculate the free energy $[F(\omega,n,\beta)]$ one has to solve 
Eqs.~\eqref{saddle-final}~\eqref{saddle-final-a}, take the analytic continuation 
$\alpha\in\mathbb{R}$, and, finally, the limit $\alpha\to 0$.  
(cf. Eq.~\eqref{replica-trick}). Although it is possible to solve 
Eqs.~\eqref{saddle-final}~\eqref{saddle-final-a} numerically for any fixed $r,
\alpha\in\mathbb{N}$, taking the analytic continuation $\alpha\in\mathbb{R}$ is a 
formidable task, since the dependence of $Z^\alpha(\omega,n,\beta)$ on $\alpha$ is in 
general too complicated. The strategy is usually to choose a specific 
form of the overlap tensor $q_{\gamma\gamma'}^{rr'}$ in terms of ``few'' 
parameters, which allows to perform the analytic continuation and the limit 
$\alpha\to 0$ exactly. 

For the standard S-K model (i.e., for $n=1$) the simplest parametrization is the 
replica-symmetric one (RS), which amounts to taking $q^{11}_{\gamma\gamma'}=q$. 
This relies on the observation that the fictitious replicas appear symmetrically 
in Eq.~\eqref{rep-Z}. Although the RS ansatz is correct at high temperatures, it 
fails in the glassy phase at low temperatures, where the permutation invariance 
within the replicas has to be broken~\cite{almeida-1978}. The celebrated Parisi 
ansatz~\cite{parisi-1979} provides a systematic scheme to break the replica 
symmetry in successive steps, and it allows to capture the glassy behavior of 
the S-K model at low temperature. We anticipate (see section~\ref{rsb-1-section} for 
the details) that a similar scheme has to be used to describe the glassy phase of 
the S-K model on the booklet.

\section{The replica-symmetric (RS) ansatz}
\label{solution}

In this section we present the solution of the S-K model on the booklet, using 
the replica symmetric (RS) approximation. In subsection~\ref{para-section} we focus 
on the high temperature phase, where this approximation is exact, and the 
behavior of the model is fully determined by the diagonal part of the overlap 
tensor $q_{\gamma\gamma'}^{rr'}$ (see section~\ref{replica-sec} for its 
definition). In~\ref{shannon} we perform the analytic continuation $n\to 1$, 
which allows to obtain the Shannon mutual information $[{\mathcal I}_1]$. 
The generic structure of $q_{\gamma\gamma'}^{rr'}$ 
within the RS ansatz  is discussed in subsection~\ref{rs-section}. This allows us 
to determine the critical temperature of the paramagnetic-glassy transition. 
For simplicity, here and in the following sections we restrict ourselves to zero 
magnetic field ($h=0$ in Eq.~\eqref{SK-ham}), and to $J_0=0$ in 
Eq.~\eqref{quenched-distr}. 

\subsection{The paramagnetic phase}
\label{para-section}

\begin{figure}[t]
\includegraphics*[width=0.9\linewidth]{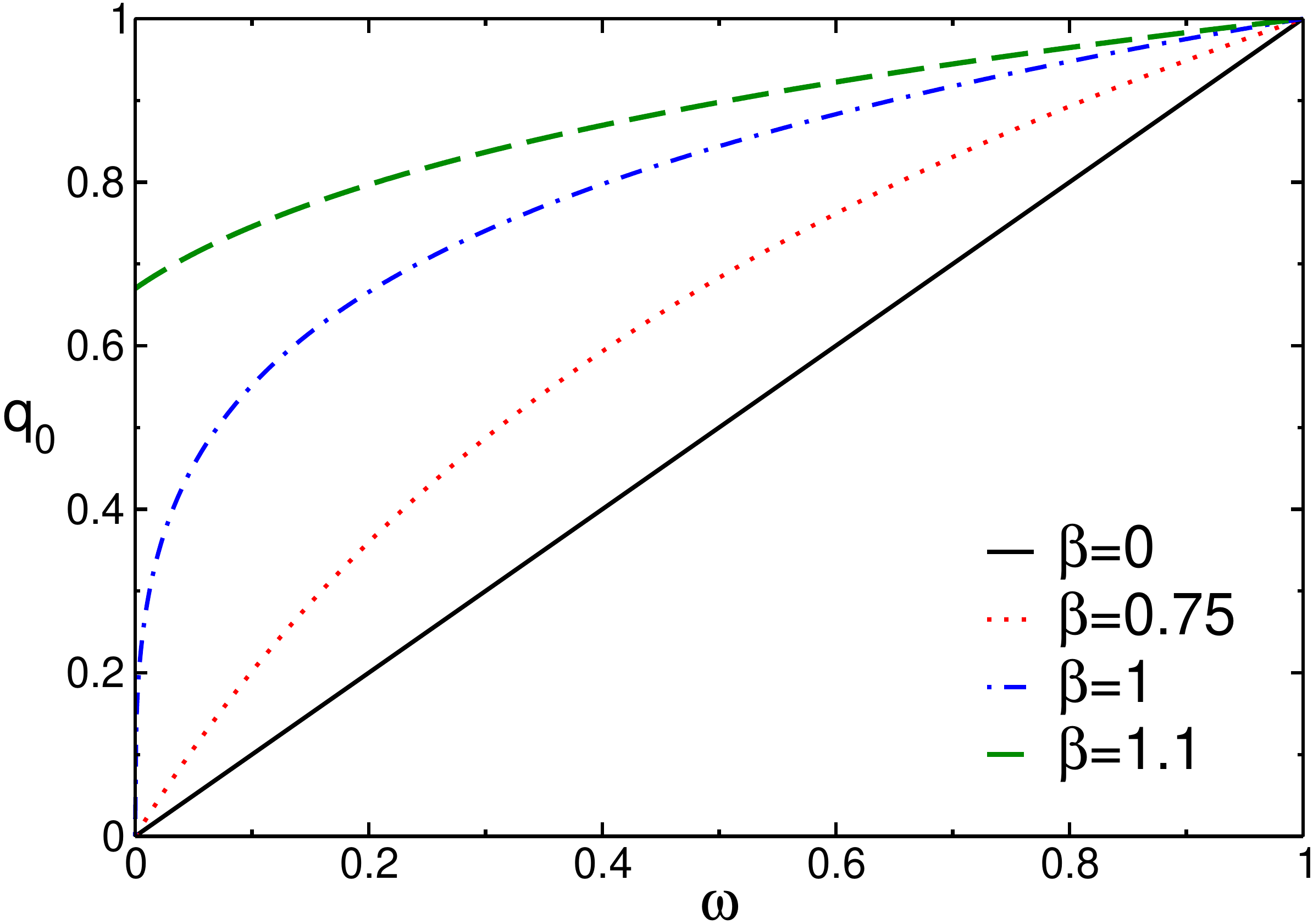}
\caption{
 The S-K model on the $2$-sheets booklet in the 
 paramagnetic phase: the solution $q_0$ of the saddle point 
 equation~\eqref{saddle-high-t} plotted as a function of the booklet aspect 
 ratio $\omega$ and several values of the inverse temperature $\beta=1/T$. 
 At $\omega=1$ one has $q_0=1,\forall\beta$. In the zero-temperature limit 
 one has $q_0\to 1,\forall\omega$. The straight line is the infinite 
 temperature result. 
}
\label{RSB0_saddle_ht}
\end{figure}

Here we provide the exact analytical expression for the disorder-averaged free 
energy $[F_{para}(\omega,n,\beta)]$  in the paramagnetic non-glassy phase. 
We start discussing the infinite temperature limit (i.e., $\beta\to 0$), 
restricting ourselves to zero magnetic field. 
Using Eq.~\eqref{mf-action}, a standard high-temperature expansion yields 
$\tr_Ae^{{\mathcal L}}=2^{\alpha}+{\mathcal O}(\beta^2)$ and $\tr_B e^{{
\mathcal L}}=2^{n\alpha}+{\mathcal O}(\beta^2)$, implying  
\begin{align}
& \langle S^{(r,\gamma)}S^{(r',\gamma')}\rangle_B=2^{n\alpha}
\delta_{\gamma,\gamma'}\delta_{r,r'}+{\mathcal O}(\beta^2)\\
& \langle S^{(r,\gamma)}S^{(r',\gamma')}\rangle_A=2^{\alpha}
\delta_{\gamma,\gamma'}+{\mathcal O}(\beta^2).
\end{align}
Using Eq.~\eqref{saddle-final}, it is straightforward to obtain the 
infinite-temperature overlap tensor $q_{\gamma\gamma'}^{rr'}$ as 
\begin{equation}
q_{\gamma\gamma'}^{rr'}=(1-\omega)\delta_{\gamma,\gamma'}\delta_{r,r'}+
\omega\delta_{\gamma,\gamma'}.
\label{inf-t-q}
\end{equation}
Notice that $q_{\gamma\gamma'}^{rr'}$ is diagonal in both the 
indices $\gamma,\gamma'$ and $r,r'$, i.e., the sheet and the  
replica spaces. Using Eq.~\eqref{Z-ac} and Eq.~\eqref{replica-trick}, 
after performing the analytic continuation $\alpha\to 0$, one obtains  
\begin{multline}
[F_{para}(\omega,n,\beta)]=N\Big\{(n-\omega(n-1))\log(2)\\
+\frac{\beta^2}{4}(\omega^2(n^2-n)+n)\Big\}+
{\mathcal O}(\beta^4).
\label{logZ-ht}
\end{multline}
It is natural to expect that for finite $\beta\le\beta_c$, with 
$\beta_c$ the critical temperature of the paramagnetic-glassy 
transition, the overlap tensor $q_{\gamma\gamma'}^{rr'}$ remains 
diagonal. This suggests the ansatz 
\begin{equation}
q_{\gamma\gamma'}^{rr'}=(1-q_0)\delta_{r,r'}\delta_{\gamma,\gamma'}+
q_0\delta_{\gamma,\gamma'}.
\label{high-t-q}
\end{equation}
with $q_0\in {\mathbb R}$ a parameter. The ansatz~\eqref{high-t-q} is formally 
obtained from Eq.~\eqref{inf-t-q} by replacing $\omega\to q_0$. Notice that 
one has $q_{\gamma\gamma}^{rr}=1$, in agreement with Eqs.~\eqref{saddle-final}. 
Using Eq.~\eqref{high-t-q}, one obtains ${\mathcal L}_{para}$ (cf. 
Eq.~\eqref{mf-action}) as   
\begin{equation}
\label{high-t-action}
{\mathcal L}_{para}=\beta^2\frac{q_0}{2}\Big\{\sum_\gamma\sum_{rr'}S^{(r,
\gamma)}S^{(r',\gamma)}-n\alpha\Big\}.
\end{equation}
After introducing the Hubbard-Stratonovich variables $z_\lambda$ (with 
$\lambda=1,\dots,\alpha$), one can write 
\begin{multline}
\label{ht-saddle}
\textrm{Tr}_B\exp({\mathcal L}_{para})=\\\tr_B\int\prod_\lambda Dz_{\lambda}
\exp\Big(z_\lambda\beta\sqrt{q_0}\sum_r S^{(r,\lambda)}\Big),
\end{multline}
where $\int Dz f(z)\equiv(2\pi)^{-1/2}\int\exp(-z^2/2)f(z)$. Notice that 
due to the square root in Eq.~\eqref{ht-saddle}, one has the constraint 
$q_0>0$. Moreover, from Eq.~\eqref{high-t-action} 
one obtains $\textrm{Tr}_A\exp({\mathcal L})=2^{n\alpha}$. The trace 
$\textrm{Tr}_B$ in Eq.~\eqref{ht-saddle} can be performed explicitly. 
Using Eq.~\eqref{Z-ac} and Eq.~\eqref{replica-trick}, one obtains 
the free energy $[F_{para}(\omega,n,\beta)]$ as  
%
\begin{figure}[t]
\includegraphics*[width=0.9\linewidth]{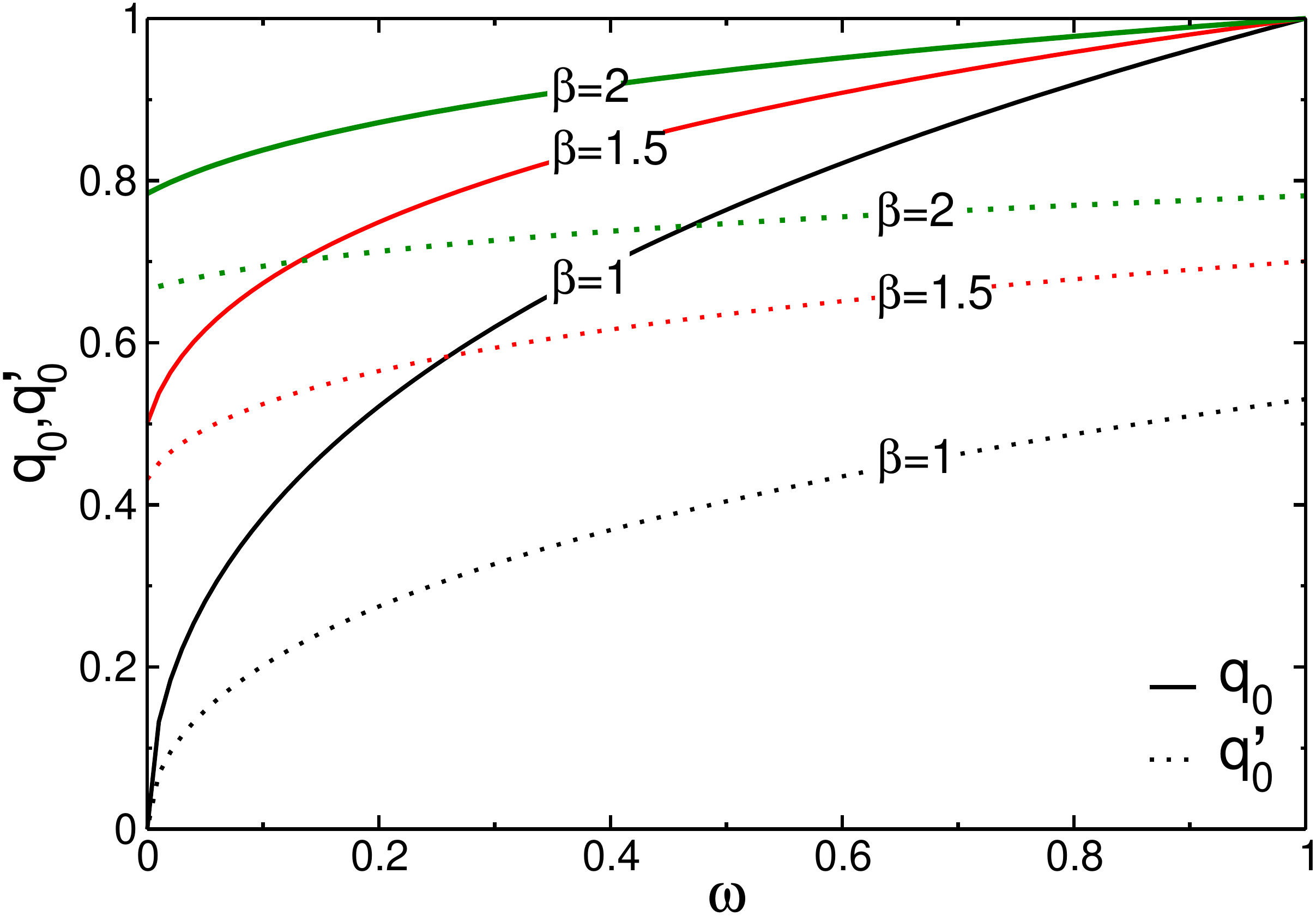}
\caption{\label{RSB0-saddle}
 The S-K model on the $2$-sheets booklet. The 
 overlap tensor in the replica-symmetric (RS) approximation (see Eq.~\eqref{rs-ansatz}): 
 the solutions $q_0$ and $q_0'$ (shown as full and dotted lines, respectively) 
 of the saddle point equations~\eqref{RS-saddle-1}\eqref{RS-saddle-2} 
 plotted as a function of the booklet aspect ratio $\omega$, and inverse 
 temperature $\beta=1/T=1,3/2,2$. At $\omega=1$ one has that $q_0=1,
 \forall\beta$. In the limit $\beta\to\infty$ it is $q_0,q_0'\to 1,\forall
 \omega$.
}
\label{RSB0_saddle}
\end{figure}
%
\begin{multline}
[F_{para}(\omega,n,\beta)]=N\Big\{
n\log(2)-\omega(n-1)\log(2)\\+
\frac{\beta^2}{2}\Big(\frac{n}{2}-\frac{q_0^2}{2}(n^2-n)
+\omega q_0n^2-q_0n\Big)\\
+(1-\omega)\log\int Dz\cosh^n(z\sqrt{q_0}\beta)
\Big\},
\label{ht-logZ}
\end{multline}
where $q_0$ is determined by solving the saddle point condition $\partial[F_{para}
(\omega,n,\beta)]/\partial q_0=0$. $[F_{para}(\omega,n,\beta)]$ can be written in 
terms of simple functions using that 
\begin{multline}
\frac{1}{\sqrt{2\pi}}\int Dz\cosh^n(z\sqrt{q_0}\beta)=\\
\frac{1}{2^{n-1}}\sum\limits_{k=0}^{\lfloor n/2\rfloor-1}\frac{\Gamma(n+1)}
{\Gamma(k+1)\Gamma(n-k+1)}e^{2\beta^2 q_0(n/2-k)^2}\\
+\frac{2^{-2\left\lfloor n/2\right\rfloor}\Gamma(n+1)}{\Gamma(n-\lfloor n/2\rfloor+1)
\Gamma(\lfloor n/2\rfloor+1)}e^{2\beta^2q_0(n/2-\lfloor n/2\rfloor)^2}, 
\end{multline}
where $\Gamma(x)$ denotes the Euler Gamma function. 
The saddle point equation for $q_0$ reads 
\begin{multline}
\label{sp-g}
0=\beta\big(\omega n-1-q_0(n-1)\big)\\
+(1-\omega)\frac{\int Dz\cosh^{n-1}(z\sqrt{q_0}\beta)\sinh(z\sqrt{q_0}\beta)z}
{\sqrt{q_0}\int Dz\cosh^n(z\sqrt{q_0}\beta)}. 
\end{multline}
For the $2$-sheets booklet (i.e., $n=2$) this is  
given as   
\begin{equation}
q_0=\omega+(1-\omega)\tanh(\beta^2 q_0).
\label{saddle-high-t}
\end{equation}
Alternatively, Eq.~\eqref{saddle-high-t} can be obtained by substituting the 
ansatz~\eqref{high-t-q} in Eqs.~\eqref{saddle-final}~\eqref{saddle-final-a}. 
Clearly, for two independent copies of the S-K model, i.e., $\omega=0$,  
Eq.~\eqref{saddle-high-t} gives $q_0=0$ for $\beta\le1$, whereas 
one has $q_0\ne0$ for $\beta>1$. On the other hand, for $\omega=1$ 
one has $q_0=1$ $\forall\beta$. For intermediate $0<\omega<1$, $q_0$ 
is plotted as a function of $\omega$ in Fig.~\ref{RSB0_saddle_ht}. For $\beta=0$ it 
is $q_0=\omega$ (straight line in the Figure). In the low-temperature limit 
one has $q_0\to 1$, for any $\omega$. In particular, it is straightforward 
to check that  $q_0\approx 1-2(1-\omega)\exp(-2\beta^2)$ for $\beta\to\infty$.

\subsection{The analytic continuation $n\to 1$}
\label{shannon}

It is interesting to consider the analytic continuation $\lim_{n\to 1} [S_n]$ 
and $\lim_{n\to 1}[{\mathcal I}_n]$ to obtain the Shannon entropy and mutual 
information. In the limit $n\to 1$, $F_{para}(\omega,n,\beta)$ (see~\eqref{ht-logZ}) 
does not depend on $q_0$ and $\omega$,  as expected. This holds for any 
$\beta$ even including the effects of the replica symmetry breaking. However, 
the Shannon entropy $S_1(A)$ (and the mutual information ${\mathcal I}_1$ thereof) 
is non trivial due to the prefactor $1/(1-n)$ in~\eqref{renyi}. In order to 
obtain $S_1(A)$ one should consider $1/(1-n)[F_{para}(\omega,n,\beta)]$ 
performing carefully the limit $n\to 1$. The result reads 
\begin{multline}
\label{sh-F}
\frac{[F_{para}(\omega,1,\beta)]}{N(1-n)}=
\omega\log(2)+
\frac{\beta^2}{2}\Big(\frac{q_0^2}{2}n
-\omega q_0(n+1)+q_0\Big)\\
-(1-\omega)\frac{\int Dz\cosh(z\sqrt{q_0}\beta)
\log\cosh(z\sqrt{q_0}\beta)}{\int Dz\cosh(z\sqrt{q_0\beta})}. 
\end{multline}
The saddle point equation~\eqref{sp-g} for $q_0$ now becomes  
\begin{multline}
(\omega-q_0)\exp\big(\beta^2\frac{q_0}{2}\big)
+(1-\omega)\int Dz\Big(\frac{z}{\beta\sqrt{q_0}}\sinh(z\sqrt{q_0}\beta)\\
-\cosh(z\sqrt{q_0}\beta)\Big)\log\cosh(z\sqrt{q_0}\beta)=0. 
\end{multline}
Notice that in the limit $\beta\to 0$ one has $q_0\to\omega$, which implies,   
using~\eqref{sh-F} and~\eqref{MI}, the volume-law behavior $[{\mathcal I}_1]
\to\beta^2/2\omega(1-\omega)N$. This in constrast with the clean 
case~\cite{wilms-2012}, where ${\mathcal I}_1={\mathcal O}(1)$ for any 
$\beta$ away from the critical point at $\beta_c=1$.

\subsection{The replica-symmetric (RS) approximation} 
\label{rs-section}

In the replica-symmetric (RS) approximation one writes the overlap tensor  
$q_{\gamma\gamma'}^{rr'}$ as
\begin{equation}
q_{\gamma\gamma'}^{rr'}=(1-q_0)\delta_{r,r'}\delta_{\gamma,\gamma'}+
q_0\delta_{\gamma,\gamma'}+(1-\delta_{\gamma,\gamma'})q'_0.
\label{rs-ansatz}
\end{equation}
The first two terms in Eq.~\eqref{rs-ansatz} are the same as in the paramagnetic 
phase (cf. Eq.~\eqref{high-t-q}). The last term sets $q_{\gamma\gamma'}^{rr'}=q'_0$ 
$\forall\gamma\ne\gamma'$ and $\forall r,r'$. Clearly, $q_{\gamma\gamma'}^{rr'}$ is 
invariant under permutations of both the booklet sheets and the replicas. We do not have any 
rigorous argument to justify the ansatz~\eqref{rs-ansatz}, besides its simplicity. 
However, we numerically observe that it captures quite accurately the behavior of 
the model, at least around the paramagnetic-glassy transition (see section~\ref{mc-results} 
for the comparison with Monte Carlo data). 

Using Eq.~\eqref{rs-ansatz} and Eqs.~\eqref{replica-trick}\eqref{Z-ac} one obtains 
the replica-symmetric approximation for the free energy $[F_{RS}(\omega,n,\beta)]$ 
as 
\begin{multline}
[F_{RS}(\omega,n,\beta)]=\\\lim_{\alpha\to 0}\Big\{
N\Big[\frac{\beta^2}{4}n-\frac{\beta^2}{4}\Big(
(q_0^2-(q'_0)^2)n^2-q_0^2n\Big)\\
+\frac{1-\omega}{\alpha}\log\tr_B\exp({\mathcal L}_{RS})\\
+\frac{\omega}{\alpha}\log\tr_A \exp({\mathcal L}_{RS})
\Big]\Big\},
\label{rs-F}
\end{multline}
where ${\mathcal L}_{RS}$ is obtained by substituting Eq.~\eqref{rs-ansatz} 
in Eq.~\eqref{mf-action}, which yields 
\begin{multline}
\label{l-rs}
{\mathcal L}_{RS}=\frac{q_0'}{2}\beta^2
\sum\limits_{\gamma\gamma'}\sum_{rr'}
S^{(r,\gamma)}S^{(r',\gamma')}\\
+\frac{q_0-q_0'}{2}\beta^2\sum_\gamma\sum_{rr'}
S^{(r,\gamma)}S^{(r',\gamma')}-\frac{q_0}{2}\beta^2n\alpha.
\end{multline}
To calculate the last two terms in Eq.~\eqref{rs-F} one has to introduce 
two auxiliary Hubbard-Stratonovich variables $z,z'$, similar to the 
paramagnetic phase (cf. section~\ref{para-section}). Thus, after performing 
the trace over the spin variables, in the limit $\alpha\to0$, one obtains 
\begin{align}
\label{eq1}
 \log\tr_B\exp({\mathcal L}_{RS})=&
-\frac{q_0}{2}\beta^2 n\alpha+n\alpha\log(2)\\\nonumber
&+\alpha\int Dz
\log\int Dz'H_{RS}^n(z,z'),\\
\label{eq2}
 \log\tr_A\exp({\mathcal L}_{RS})=&
-\frac{q_0}{2}\beta^2 n\alpha+\alpha\log(2)\\\nonumber
&+\alpha\int Dz
\log\int Dz'H_{RS}(nz,nz'), 
\end{align}
where $H_{RS}(z,z')\equiv\cosh(\beta z\sqrt{q_0'}+\beta z'\sqrt{q_0-q_0'})$. 
Notice that because of the square roots in the definition of $H_{RS}(z,z')$, 
one has the constraint $0\le q_0'\le q_0\le 1$. 

In Eqs.~\eqref{eq1}\eqref{eq2} $q_0,q_0'$ satisfy the saddle point conditions 
$\partial[F_{RS}(\omega,n,\beta)]/\partial q_0=\partial[F_{RS}(\omega,n,\beta)]/
\partial q'_0=0$ (cf. Eqs.~\eqref{RS-saddle-1}\eqref{RS-saddle-2} for their 
form for $n=2$). The resulting $q_0$ and $q_0'$ are plotted in Fig.~\ref{RSB0-saddle} 
(full and dotted lines, respectively) as function of $\omega$ and for several 
values of $\beta$. Clearly, for any $\beta$ one has $q_0=1$ in the limit 
$\omega\to 1$. Also, in the zero-temperature limit $\beta\to\infty$ one has 
that $q_0\to 1$ and $q_0'\to 1$, for any $\omega$. Moreover, a simple large $\beta$ 
expansion yields  
\begin{align}
\label{low-t}
& q_0=1 -(1-\omega)\sqrt{\frac{2}{\pi}}\frac{1}{\beta}\exp\Big(-\frac{1}{\pi}-
\sqrt{\frac{2}{\pi}}\beta\Big)+\dots
\\ 
& q_0'=1-\frac{1}{\sqrt{2\pi}\beta}-\frac{1}{2\pi\beta^2}+{\mathcal O}(\beta^{-3}),
\end{align}
with the dots denoting exponentially suppressed terms in the limit $\beta\to\infty$. 
Interestingly, from Eq.~\eqref{low-t} one has that $q_0\to 1$ exponentially in the limit $\beta\to\infty$, 
as in the paramagnetic phase (cf. section~\ref{para-section}), whereas 
$q_0'-1\propto 1/\beta$.

\begin{figure}[t]
\includegraphics*[width=0.9\linewidth]{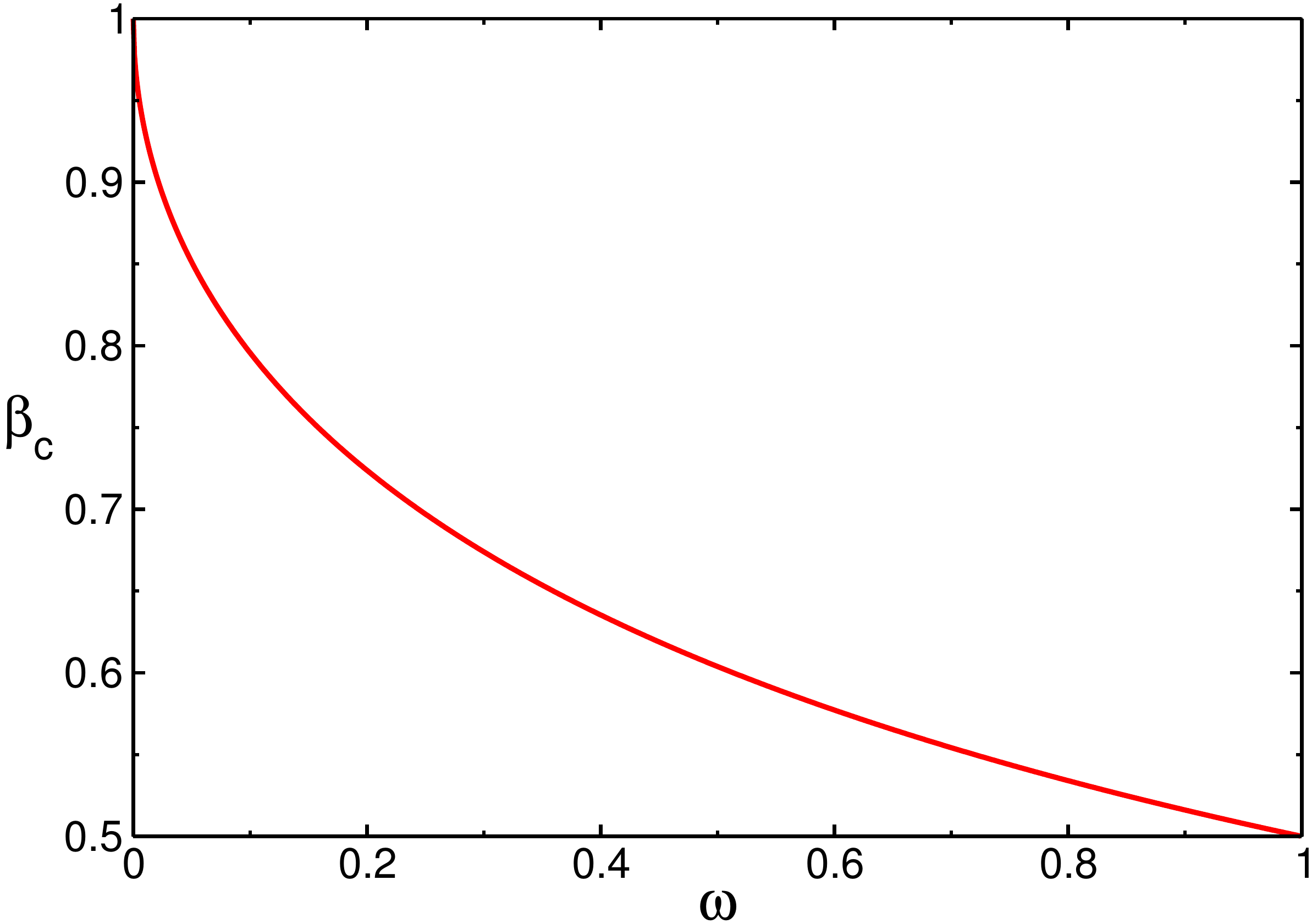}
\caption{
 The critical temperature $\beta_c\equiv 1/T_c$ of the paramagnetic-glassy 
 transition for the S-K model on the $2$-sheets booklet (see Fig.~\ref{cartoon}) 
 from Eq.~\eqref{tc}: $\beta_c$ as a function of the booklet ratio $\omega
 \equiv N_A/N$. Here $\beta_c$  is obtained from the replica-symmetric (RS) 
 approximation. Notice that $\beta_c=1$ and $\beta_c=1/2$ for $\omega=0$ and 
 $\omega=1$, respectively. 
}
\label{beta_c}
\end{figure}

\subsection{The paramagnetic-glassy transition}
\label{tc-section}

Using the replica-symmetric ansatz Eq.~\eqref{rs-ansatz} one can determine 
the critical temperature $\beta_c$ of the paramagnetic-glassy transition. 
Near the glassy transition one should expect $q'_0\to0$, whereas $q_0$ 
should remain finite (see section~\ref{para-section}). 
One expands $[F_{RS}(\omega,n,\beta)]$ 
(cf. Eq~\eqref{rs-F}) for small $q_0'$, keeping only terms up to ${\mathcal O}((q_0')^2)$. 
Thus, $\beta_c$ is obtained by imposing that the coefficient of the quadratic term $q_0'^2$ 
vanishes. This leads to the equation 
%
\begin{equation}
\exp(-4q_0\beta^2_c)+2\exp(-2q_0\beta^2_c)=\frac{1-4\beta^2_c}
{4\beta^2_c\omega-1},
\label{tc}
\end{equation}
where $q_0$ is obtained by solving the high-temperature saddle point equation~\eqref{saddle-high-t}. 
The resulting $\beta_c$ is plotted in Fig.~\ref{beta_c} as a function of $\omega$.

\section{The one-step replica-symmetry-breaking (1-RSB) approximation}
\label{rsb-1-section}

In this section we go beyond the replica-symmetric approximation, including some 
of the effects of the replica symmetry breaking. More specifically, here we discuss 
the one-step replica symmetry breaking (1-RSB) approximation. The overlap tensor 
$q_{\gamma\gamma'}^{rr'}$ now reads  
\begin{equation}
\label{q-rsb1}
q_{\gamma\gamma'}^{rr'}=(1-q_0)\delta_{\gamma,\gamma'}\delta_{r,r'} +
q_0\delta_{\gamma,\gamma'}+(1-\delta_{\gamma,\gamma'})q', 
\end{equation}
which is formally equivalent to the RS ansatz in Eq.~\eqref{rs-ansatz}, apart from 
the trivial redefinition $q_0'\to q'$. However, in contrast with Eq.~\eqref{rs-ansatz}, 
where $q_0'\in\mathbb{R}$ is a number, here $q'$ is a matrix. Inspired by the Parisi 
scheme for the standard S-K model~\cite{parisi-1979}, we choose 
\begin{equation}
q'=\left\{
\begin{array}{cc}
q_1' & \textrm{if}\, \lfloor\gamma/m_1\rfloor=\lfloor\gamma'/m_1\rfloor\\\\
q_0' & \textrm{otherwise}\\
\end{array}
\right.
\label{q-rsb1a}
\end{equation}
where $q_0',q_1'\in{\mathbb R}$, $m_1\in{\mathbb N}$, and $\lfloor\cdot\rfloor$ 
denotes the floor function. Notice that the off-diagonal elements of $q'$ (i.e., for 
$\gamma\ne\gamma'$) do not depend on $r,r'$, meaning that, although the permutation 
symmetry between the replicas is broken, the symmetry among the booklet sheets  
is preserved. The choice in Eq.~\eqref{q-rsb1a} corresponds to a simple block-diagonal 
structure for $q'$: the matrix elements of the $m_1\times m_1$ diagonal blocks of $q'$ 
are set to $q_1'$, whereas all the off-diagonal elements are set to $q_0'$. As for the 
replica-symmetric ansatz in Eq.~\eqref{rs-ansatz}, we do not have any rigorous argument 
to justify Eq.~\eqref{q-rsb1a} (see section~\eqref{mc-results}, however, for numerical 
results).

The effective interaction ${\mathcal L}_{1\textrm{-}RSB}$ (cf. Eq.~\eqref{mf-action}) 
in the replica space is obtained by substituting Eq.~\eqref{q-rsb1} in Eq.~\eqref{mf-action}. 
This yields  
\begin{multline}
{\mathcal L}_{1\textrm{-}RSB}/\beta^2=-\frac{q_0}{2}n\alpha
-\frac{q_1'-q_0}{2}\sum\limits_{\sigma=1}^{\alpha/m_1}
\sum\limits_{\gamma\in B_\sigma}\Big(\sum_r S^{(r,\gamma)}\Big)^2\\
+\frac{q_0'}{2}\Big(\sum\limits_{\gamma,r}
S^{(r,\gamma)}\Big)^2-
\frac{q_0'-q_1'}{2}\sum\limits_{\sigma=1}^{\alpha/m_1}
\Big(\sum\limits_{\gamma\in B_\sigma,r}S^{(r,\gamma)}\Big)^2,
\label{blocks}
\end{multline}
where we defined $B_\sigma\equiv[\sigma m_1,(\sigma+1)m_1)$ with $\sigma\in\mathbb{N}$.

\begin{figure}[t]
\includegraphics*[width=0.93\linewidth]{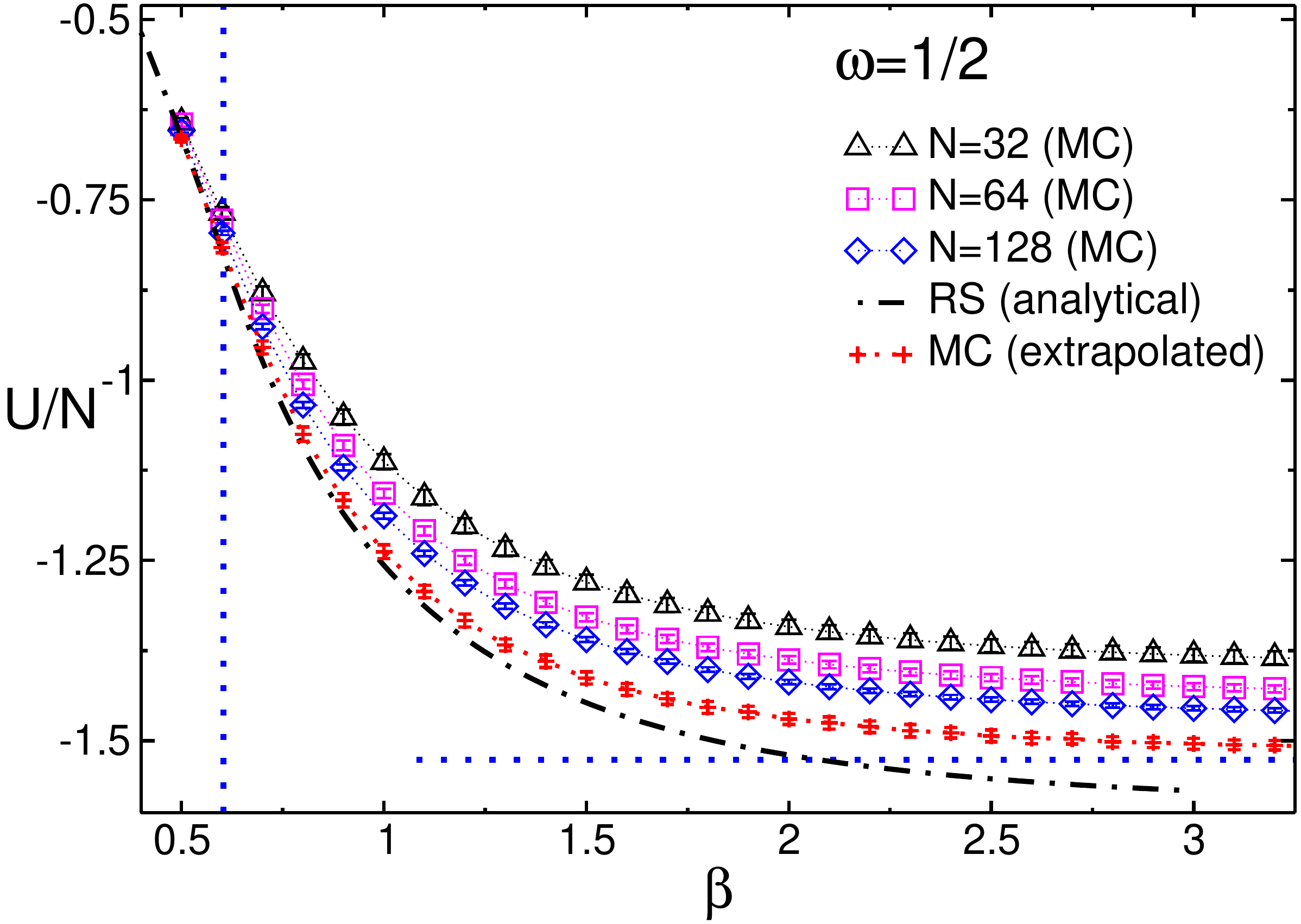}
\caption{
 The S-K model on the $2$-sheets booklet with aspect ratio $\omega=1/2$ 
 (see Fig.~\ref{cartoon}): The internal energy 
 per spin $U/N$ as a function of the inverse temperature $\beta$. The triangles, 
 squares, and rhombi, denote the Monte Carlo data for a booklet with 
 $N=32,64,128$ spins per sheet. Notice that the Monte Carlo error bars are 
 often smaller than the symbol sizes. The plus symbols are the extrapolations to the 
 thermodynamic limit $N\to\infty$, at fixed $\omega$. The dash-dotted line is the 
 analytical result $U_{RS}/N$ obtained using the replica-symmetric (RS) approximation. 
 The horizontal dotted line is the exact zero-temperature result $U/N=2u_{\infty}$, 
 with $\tilde u_{\infty}\approx-0.76321$ the zero temperature energy density of the 
 regular S-K model~\cite{parisi-1979,parisi-1983}.
}
\label{U-MC}
\end{figure}

It is convenient to introduce the Hubbard-Stratonovich variables $z,z_\sigma,z_{\sigma,\gamma}$ 
(one for each term in Eq.~\eqref{blocks}). One then obtains 
\begin{multline}
\label{step}
\log\tr'\exp({\mathcal L}_{1\textrm{-}RSB})=
-\frac{q_0}{2}\beta^2 n\alpha\\
+\log\tr'\int Dz\prod_\sigma\int Dz_\sigma\prod\limits_{\gamma\in B_\sigma}
\int Dz_{\sigma,\gamma}\prod_r\exp\Big\{\\
\beta\Big(z\sqrt{q'_0}+z_{\sigma,\gamma}\sqrt{q_0-q_1'}+
z_{\sigma}\sqrt{q_1'-q_0'}\Big)S^{(r,\gamma)}\Big\}.
\end{multline}
The trace over the spin variables in Eq.~\eqref{step} can be now performed explicitly. 
Finally, one obtains the 1-RSB approximation for the free energy 
$[F_{1\textrm{-}RSB}(\omega,n,\beta)]$ as 
\begin{widetext}
\begin{multline}
[F_{1\textrm{-}RSB}(\omega,n\beta)]/N=
(\omega+(1-\omega)n)\log(2)+
\frac{n}{4}\beta^2\Big(1+nq_0'^2
-n(m_1-1)(q_1'^2-q_0'^2)-
(n-1)q_0^2-2q_0\Big)\\
+\int Dz\left\{\frac{1-\omega}{m_1}\log\int Dz'\Big\{\int Dz''
H_{1\textrm{-}RSB}^n(z,z',z'')\Big\}^{m_1}
+\frac{\omega}{m_1}
\log\int Dz'\Big\{\int Dz''
H_{1\textrm{-}RSB}(nz,nz',nz'')\Big\}^{m_1}
\right\}, 
\label{RSB-1-logZ}
\end{multline}
\end{widetext}
where we defined $H_{1\textrm{-}RSB}(z,z',z'')$ as 
\begin{multline}
H_{1\textrm{-}RSB}(z,z',z'')\equiv\cosh(z\beta\sqrt{q'_0}\\+
z''\beta\sqrt{q_0-q'_1} +z'\beta\sqrt{q_1'-q_0'}).
\label{H1}
\end{multline}
Similar to the replica-symmetric situation (see section~\ref{rs-section}), 
from Eq.~\eqref{H1} one has the constraint $0\le q_0'\le q_1'\le q_0\le 1$.
The parameters $q_0,q_0',q_1',m_1$ are obtained by solving the saddle 
point equations~\eqref{RSB-1-saddle-a}\eqref{RSB-1-saddle-b}\eqref{RSB-1-saddle-c}
\eqref{RSB-1-saddle-d}. One should remark that, although $m_1$ is by defintion an 
integer, one obtains $m_1\in\mathbb{R}$ from the saddle point equations. 
Clearly, the replica-symmetric result $[F_{RS}(\omega,n,\beta)]$ (cf. 
Eq.~\eqref{rs-F}) is recovered from Eq.~\eqref{RSB-1-logZ} in the limit 
$q_1'=q_0'$, while the free energy in the paramagnetic phase $[F_{para}(\omega,n,
\beta)]$ (cf. Eq.~\eqref{ht-logZ}) corresponds to $q_1'=q_0'=0$.

\section{Monte Carlo results: The internal energy}
\label{mc-results}

In this section we numerically confirm the analytical results of section~\ref{solution}. 
We discuss Monte Carlo (MC) data for the S-K model on the $2$-sheets booklet 
with zero external magnetic field. The data we present are obtained from parallel tempering  
Monte Carlo simulations~\cite{hukushima-1996}. In the parallel tempering simulations 
$N_{\textrm{rep}}$ identical disorder realizations of the system are simulated. Each copy is at 
a different temperature in the range $\beta_{\textrm{min}}-\beta_{\textrm{max}}$. 
The standard Metropolis sweeps at 
each temperature are supplemented with parallel tempering moves, which allow 
to exchange the spin configurations between replicas with neighboring temperatures. 
In our simulations we choose the interval with $\beta_{\textrm{min}}=0.25$ and 
$\beta_{\textrm{max}}=2.5$, with $46$ temperatures corresponding to equally spaced 
values of $\beta$. Each Monte Carlo sweep consists of $N$ single spin updates and $45$ 
parallel tempering moves. For each disorder realization we perform $10^5$ sweeps. 
The disorder average is performed over $128$ different disorder realizations.  

We focus on the internal energy $U(\omega,n,
\beta)$ 
\begin{equation}
\label{U-def}
U(\omega,n,\beta)\equiv-\frac{\partial}{\partial\beta}
[\log(Z(\omega,n,\beta))]. 
\end{equation}
Fig.~\ref{U-MC} plots the MC data for $U(\omega,2,\beta)$ versus the inverse temperature 
$\beta$, for $\omega=1/2$. The circles, squares, and triangles are the MC results for 
different sizes, i.e., number of spins per sheet, $N=32,64,128$. The vertical dotted line 
is the critical temperature $\beta_c\approx 0.6$ of the paramagnetic-glassy transition 
(cf. Fig.~\ref{beta_c}). In the high-temperature region finite-size effects are small, 
and already for $N=64$ the MC data are indistinguishable from the thermodynamic limit 
result. Oppositely, stronger scaling corrections are visible in the low-temperature phase at 
$\beta>\beta_c$. The plus symbols in Fig.~\ref{U-MC} are the numerical extrapolations in 
the thermodynamic limit. These are obtained by fitting the finite size MC data 
to the ansatz $U/N=u_{\infty}(\omega,\beta)+c/N^{\phi}$, where 
$u_{\infty}$ is energy density in the thermodynamic limit, $c$ a fitting parameter, and 
$\phi$ the exponent of the scaling corrections. In our fits we fix $\phi=2/3$, which 
is the exponent governing the finite-size corrections of $U/N$ in 
the standard S-K model~\cite{billoire-2007,aspelmeier-2008}. 

The dash-dotted line in Fig.~\ref{U-MC} is the analytical result $U_{RS}$ obtained using 
the replica-symmetric (RS) approximation (see section~\ref{rs-section}). 
Using Eq.~\eqref{rs-F} and Eq.~\eqref{U-def}, $U_{RS}$  is obtained as   
\begin{equation}
U_{RS}=-N\beta(1+q_0^2-2q_0'^2). 
\label{U}
\end{equation}
Here $q_0,q_0'$ are solutions of the saddle point equations Eqs.~\eqref{RS-saddle-1}
\eqref{RS-saddle-2}. Notice that $U_{RS}$ depends on $\omega$ only through $q_0,q_0'$. 
From Fig.~\ref{U-MC} one has that, while $U_{RS}$ is in perfect agreement with the 
numerics for $\beta\approx\beta_c$, deviations appear in the low-temperature region. 
Notice that already at $\beta\gtrsim 1$, $U_{RS}$ is incompatible with the data. 
These deviations increase upon lowering the temperature and  have  to be attributed 
to the replica symmetry breaking happening in the glassy phase. Finally, since in 
the limit $\beta\to\infty$ all the sheets  are in the same state, 
one should expect that $u_\infty(\omega,\beta)\to n \tilde u_\infty$ (horizontal line in 
Fig.~\ref{U-MC}), with $\tilde u_{\infty}=-0.76321...$~\cite{parisi-1979,parisi-1983} the 
zero-temperature internal energy density of the S-K model on the plane.

\begin{figure}[t]
\includegraphics*[width=0.93\linewidth]{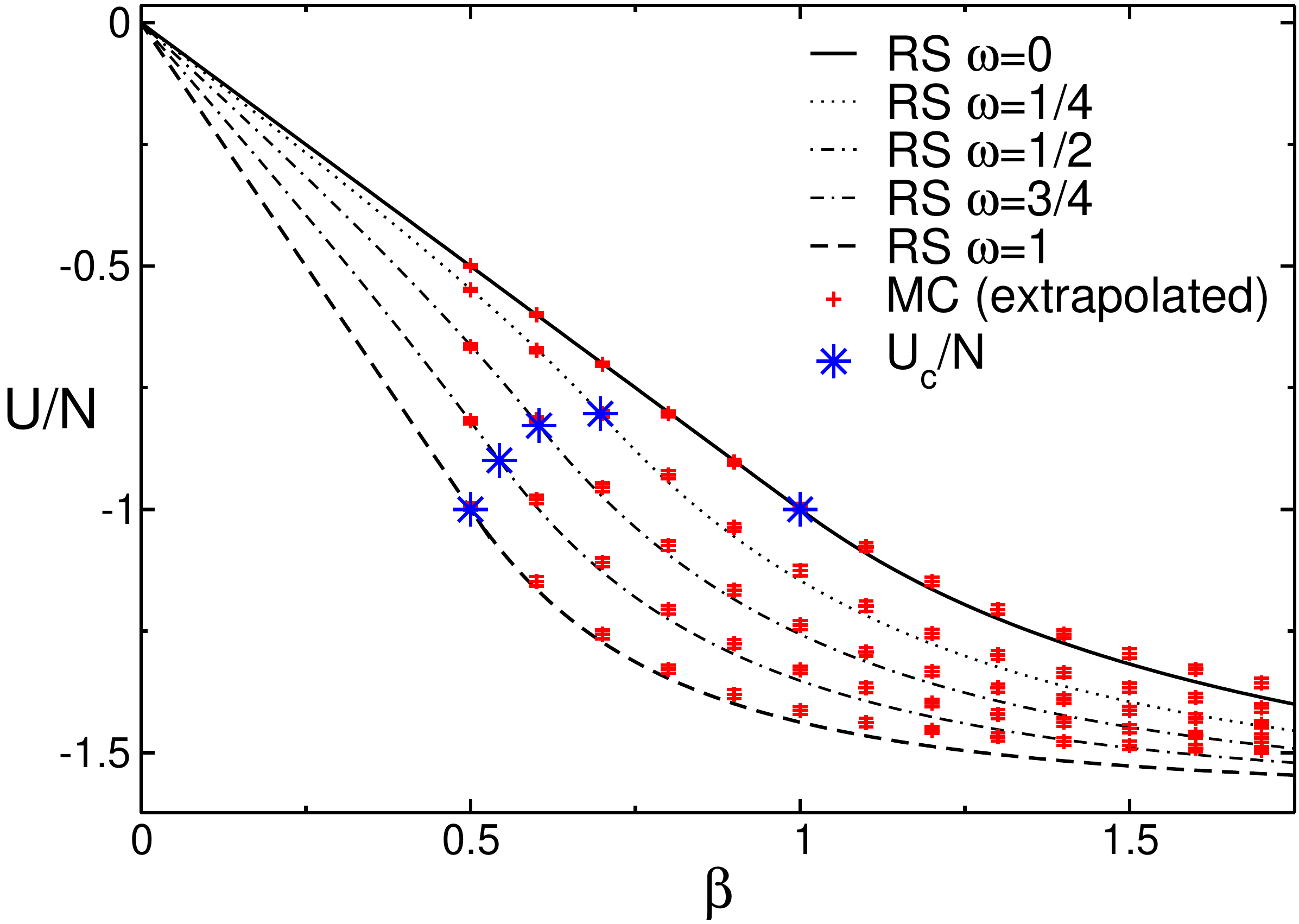}
\caption{
 The S-K model on the 2-sheets booklet: 
 The internal energy per spin $U/N$ in the thermodynamic limit. 
 The symbols are the Monte Carlo data extrapolated to the thermodynamic 
 limit $N\to\infty$, at fixed booklet ratio $\omega$ (see Fig.~\ref{cartoon}).  
 $U/N$ is plotted as a function of inverse temperature $\beta$ and for 
 $\omega=0,1/4,1/2,3/4,1$. The lines are the analytical results $U_{RS}$ 
 obtained using the replica-symmetric (RS) approximation. The stars denote 
 the value of $U_c/N$ at the paramagnetic-glassy transition. 
}
\label{U-RS}
\end{figure}

\begin{figure}[t]
\includegraphics*[width=0.93\linewidth]{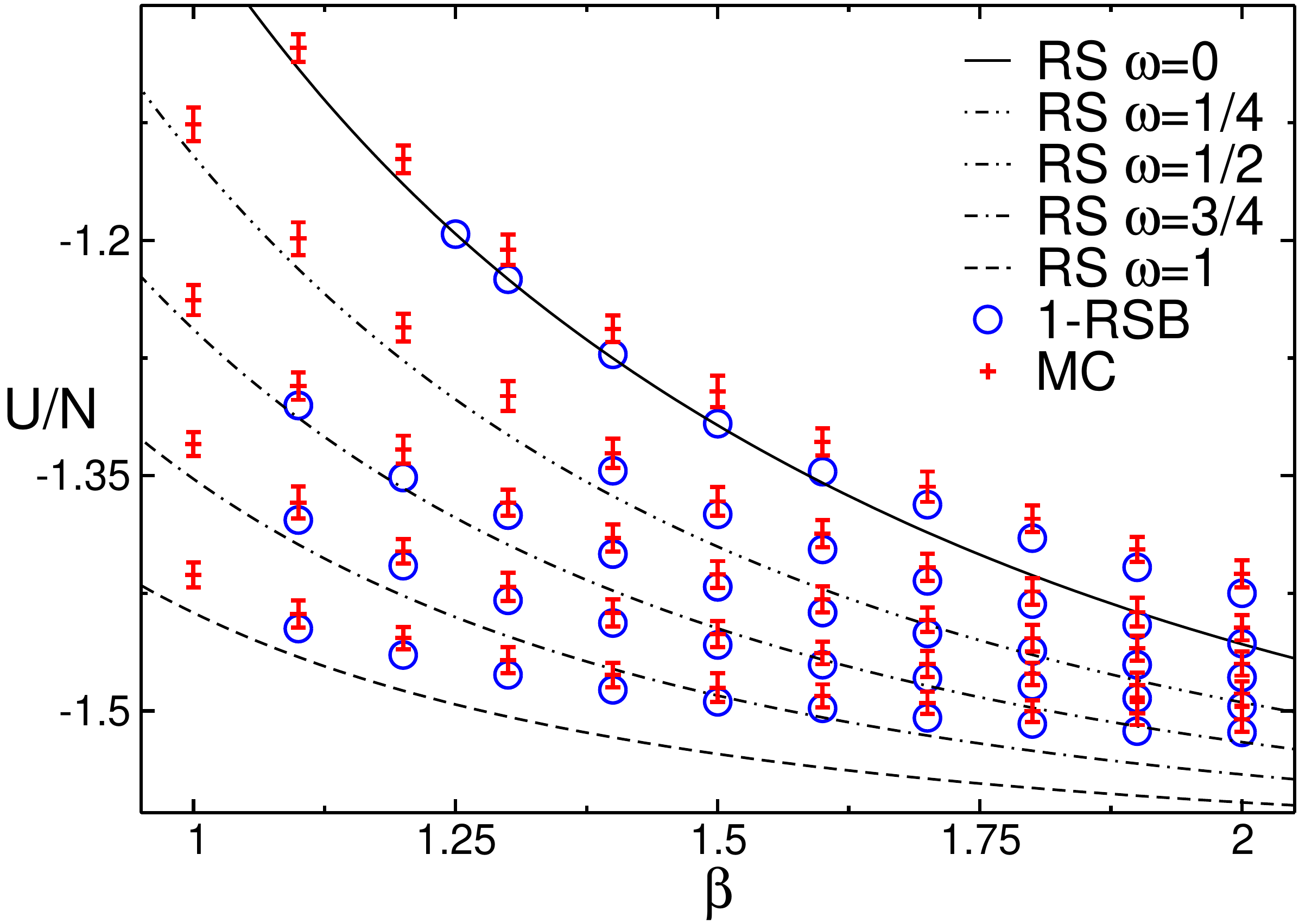}
\caption{
 The S-K model on the 2-sheets booklet: 
 The internal energy per spin $U/N$ in the thermodynamic limit plotted 
 as a function of inverse temperature $\beta$. The plus symbols are the same 
 extrapolated Monte Carlo data as in Fig.~\ref{U-RS}. The lines denote $U/N$ 
 in the replica-symmetric (RS) approximation (same as in Fig.~\ref{U-RS}). 
 The circles are the results  in the one-step replica symmetry breaking 
 (1-RSB) approximation.
}
\label{U-RSB-1}
\end{figure}

\begin{figure*}[t]
\includegraphics*[width=0.93\linewidth]{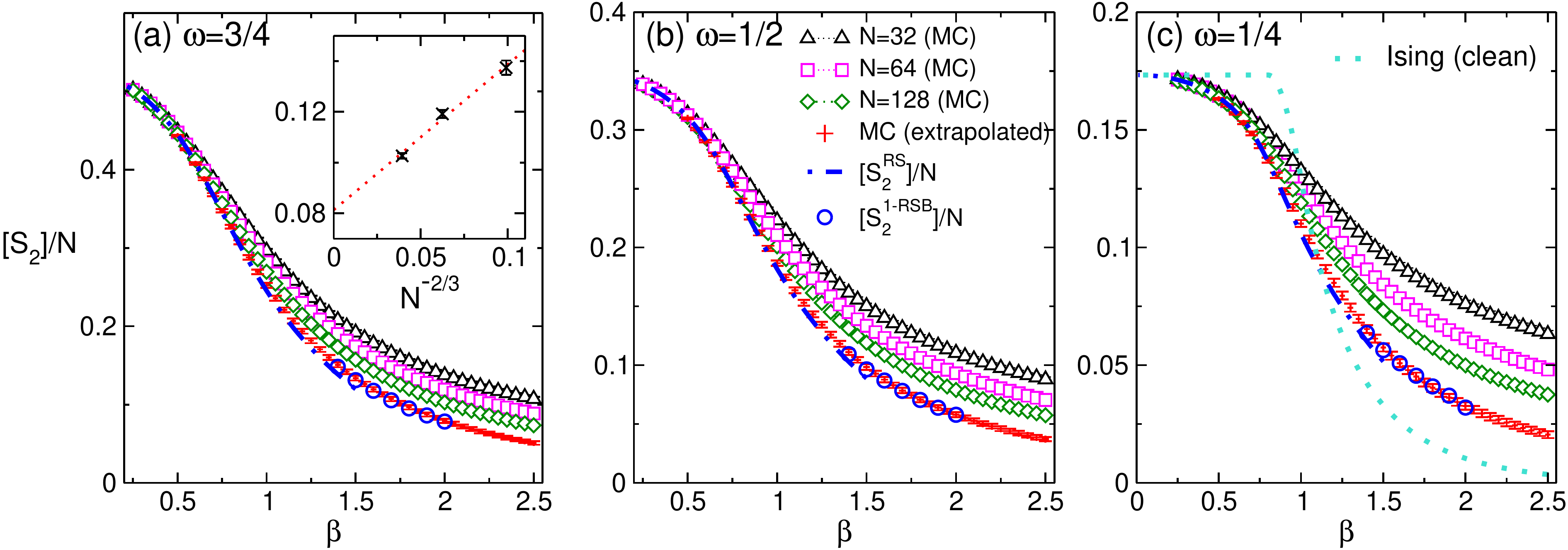}
\caption{ The classical disorder-averaged R\'enyi entropy per spin $[S_2(\omega)]/N$  
 in the S-K model on the $2$-sheets booklet: 
 $[S_2(\omega)]/N$ as a function of the inverse temperature $\beta$. The different panels 
 correspond to different booklet aspect ratios $\omega=3/4,1/2,1/4$ (see Fig.~\ref{cartoon}). 
 The triangles, squares and rhombi denote the Monte Carlo results for 
 booklets with $N=32,64,128$ spins per sheet. The Monte Carlo error bars are often 
 smaller than the symbol size. The plus symbols are the numerical 
 extrapolations to the thermodynamic limit. The inset in panel (a) shows 
 $[S_2]/N$ at fixed $\beta=2$, plotted versus $N^{-2/3}$.The dotted line is 
 a linear fit. In all the panels  the dash-dotted lines are the  
 analytical results in the replica-symmetry (RS) approximation. 
 The circles are the analytical results obtained  using the one-step replica 
 symmetry breaking (1-RSB) ansatz. In panel (c) the dotted line (clean Ising) is the 
 analytic result for the infinite-range Ising model without disorder. 
}
\label{Renyi-MC}
\end{figure*}

The behavior of $U/N$ for different $\omega$ is investigated in Fig.~\ref{U-RS},
plotting $U/N$ as a function of $\beta$ and for $\omega=0,1/4,3/4,1$. The result for $\omega=1/2$ 
is shown for comparison. The plus symbols 
are the MC data extrapolated to the thermodynamic, at fixed $\omega$. Similar to Fig.~\ref{U-RS}, 
the extrapolations are done assuming $U/N(\omega,n,\beta)=u_{\infty}(\omega,n,\beta)+
c/N^{\phi}$, with $\phi=2/3$ irrespective of $\omega$. The stars in Fig.~\ref{U-RS} are 
the critical values $U_c/N$ at the paramagnetic-glassy transition, whereas the lines 
are the analytical results (cf. Eq.~\eqref{U}). Notice that at high temperature, and for 
generic $n$ and $\omega$, Eq.~\eqref{logZ-ht} and Eq.~\eqref{U-def} give  
\begin{equation}
U_{RS}= -N\frac{\beta}{2}(\omega^2(n^2-n)+n)+{\mathcal O}(\beta^2), 
\label{ht-U}
\end{equation}
i.e., a linear behavior of $U/N$ as a function of $\beta$. For $\omega=0$ and $\omega=1$ 
this behavior is exact up to the critical point at $\beta=\beta_c$, meaning that the higher 
orders ${\mathcal O}(\beta^2)$ in Eq.~\eqref{ht-U} are zero. This is only an approximation 
at intermediate $0<\omega<1$. Both the behaviors in Eq.~\eqref{U} and Eq.~\eqref{ht-U} are 
clearly confirmed in Fig.~\ref{U-RS}. 

However, from Fig.~\ref{U-RS} one has that the RS result is not correct for $\beta>\beta_c$, 
where the replica symmetry breaking has to be taken into account. This is  more carefully 
discussed in Fig.~\ref{U-RSB-1}, focusing on the low temperature region at $1\le\beta\le 2$. 
The plus symbols and the lines are the same as in Fig.~\ref{U-RS}. The circles denote the 
internal energy per spin $U_{1\textrm{-}RSB}/N$ as obtained using the $1$-step replica 
symmetry breaking (1-RSB) approximation (see section~\ref{rsb-1-section}). Specifically, 
from Eq.~\eqref{RSB-1-logZ} and Eq.~\eqref{U-def}, for $n=2$ a straightforward calculation 
gives 
\begin{equation}
U_{1\textrm{-}RSB}=-N\beta(1+q_0^2+2q_1'^2(m_1-1)-2q_0'^2m_1), 
\label{U'}
\end{equation}
where $q_0,q_0',q_1',m_1\in\mathbb{R}$ are solutions of the saddle point 
equations \eqref{RSB-1-saddle-a}-\eqref{RSB-1-saddle-d}. Clearly, Eq.~\eqref{U'} 
implies $U_{1\textrm{-}RSB}\to U_{RS}$ for $q_1'\to q_0'$, as expected. 
Moreover, for $\beta\approx\beta_c$ one has $q_1'\approx q_0'\approx0$, implying   
that $U_{1\textrm{-}RSB}\approx U_{RS}$, i.e. the effects of the replica symmetry 
breaking are negligible near the critical point. Interestingly, at low temperatures, where the 
RS approximation fails (see Fig.~\ref{U-RS}), $U_{1\textrm{-}RSB}$ 
is in good agreement with the Monte Carlo data, at least up to $\beta\approx 2$.

\section{The classical R\'enyi entropies}
\label{Renyi-section}

We now turn to discuss the behavior of the classical R\'enyi entropies 
(cf. Eq.~\eqref{renyi}). Here we restrict ourselves to the second R\'enyi 
entropy $[S_2]$, which is obtained in terms of the booklet partition function 
$Z(\omega,2,\beta)$ (see section~\ref{booklet}) as $-\log(Z(\omega,2,\beta)/Z(0,2,\beta))$. 
Due to the mean-field nature of the S-K model (cf. Eq.~\eqref{SK-ham}), 
there is no well defined boundary between the two parts $A$ and $B$ of the system 
(unlike in local spin models). With this in mind we assume that we will find volume law 
behavior $[S_2]\propto N$, and consider the entropy per spin 
$[S_2]/N$ as the quantity of interest. 

\begin{figure*}[t]
\includegraphics*[width=0.93\linewidth]{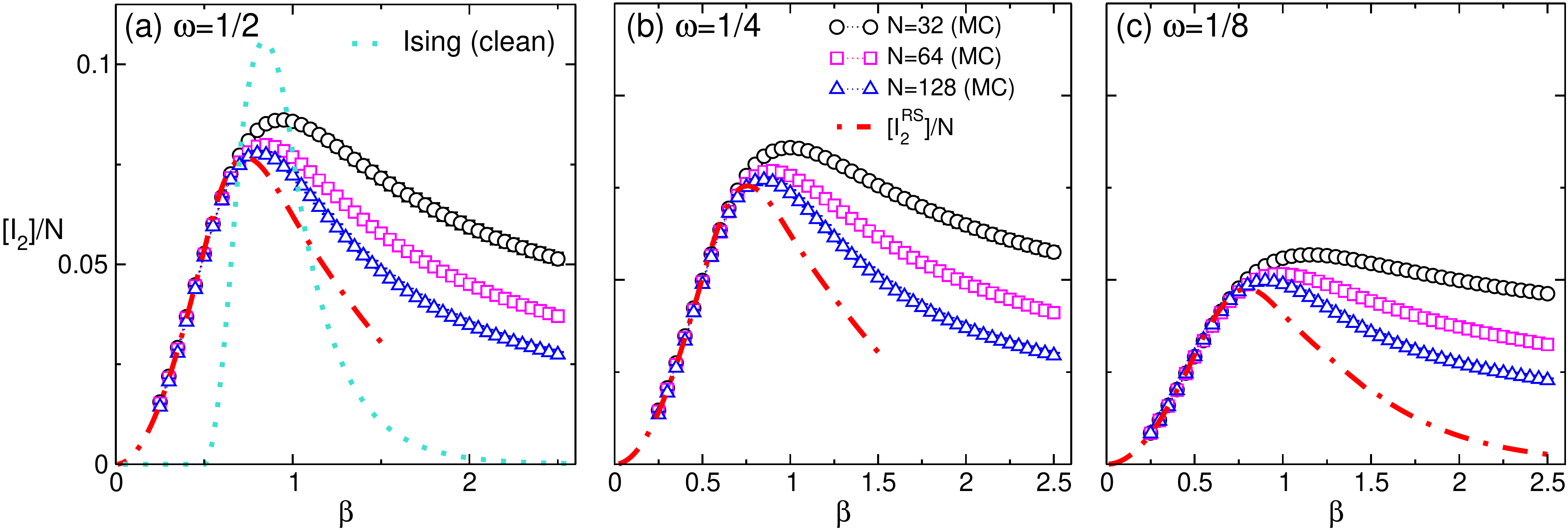}
\caption{The classical disorder-averaged mutual information per spin $[{\mathcal I}_2]/N$ 
 in the S-K model on the $2$-sheets booklet: 
 $[{\mathcal I}_2]/N$ versus the inverse temperature $\beta$. The 
 different panels correspond to different booklet ratios $\omega=1/2,1/4,1/8$ 
 (see Fig.~\ref{cartoon}). The same scale is used on both axes in all 
 panels. The symbols are the Monte Carlo data for systems with $N=32,64,128$ 
 spins per sheet. The Monte Carlo error bars are smaller than the symbol size. 
 The dash-dotted line is the analytical result in the replica-symmetric (RS) 
 approximation. In (a) the dotted line is the result for the clean, i.e., without 
 disorder, infinite-range Ising model. 
}
\label{I2-MC}
\end{figure*}

The Monte Carlo data for $[S_2]/N$  are shown in Fig.~\ref{Renyi-MC} plotted versus the 
inverse temperature $0.25\le\beta\le 2.5$. Some details on the Monte Carlo method used 
to calculate $[S_2]$ are provided in Appendix~\ref{mc-method}. The different panels 
correspond to the booklet ratios $\omega=3/4,1/2,1/4$ (see Fig.~\ref{cartoon}). In all the 
panels the triangles, squares, and rhombi correspond to booklets with $N=32,64,128$ spins 
per sheet. Clearly, finite size effects are present, which increase upon lowering 
the temperature, as expected. In order to obtain $[S_2]/N$ in the thermodynamic limit 
we fit the data to the ansatz $[S_2]/N=s_2(\omega)+c'(\omega)/N^{\phi}$, where 
$s_2(\omega)$ is the entropy per spin in the thermodynamic limit, $c'$ a constant, and 
$\phi$ the exponent of the finite-size corrections. The plus symbols in Fig.~\ref{Renyi-MC} 
are the results of the fits. We should mention that the fits give $\phi\approx 2/3$, 
which is the exponent of the scaling corrections of the free energy in the standard 
S-K model. This is not surprising, since $[S_2]$ is obtained as the difference 
$[S_2]\equiv [F(0,2,\beta)-F(\omega,2,\beta)]$ (cf. Eq.~\eqref{renyi}). 
Clearly, from Fig.~\ref{Renyi-MC} one has that in the thermodynamic limit $[S_2]/N$ is 
finite for any $\beta$, confirming the expected volume law behavior. Moreover, $[S_2]/N$ 
exhibits a maximum in the infinite-temperature limit $\beta\to0$. The height of this 
maximum  is a decreasing function of $\omega$ (compare the panels (a)(b)(c) in 
Fig.~\ref{Renyi-MC}). 

The dash-dotted line in the Figure denotes the analytical result $[S_2^{RS}]$ obtained within 
the RS approximation (see section~\ref{rs-section}). More precisely, $[S_2^{RS}]/N$ is obtained 
from Eq.~\eqref{renyi} and the expression for the free energy $[F_{RS}]$ (cf. Eq.~\eqref{rs-F}). 
Notice that in the high-temperature limit $\beta\to0$, where the RS approximation is exact, 
Eq.~\eqref{renyi} and Eq.~\eqref{ht-logZ} give 
\begin{equation}
\frac{[S^{RS}_2]}{N}=\omega\log(2)-\frac{\beta^2}{4}\omega^2+{\mathcal O}(\beta^3).
\end{equation}
From Fig.~\ref{Renyi-MC} one has that the extrapolated MC data are in quantitative 
agreement with $[S_2^{RS}]/N$ for $\beta\lesssim 1.5$, whereas strong deviations are 
observed at lower temperatures (not shown in the Figure). A better approximation for 
$[S_2]/N$ at low temperatures is obtained by including the effects of the replica 
symmetry breaking. The circles in Fig.~\ref{Renyi-MC} denote the one-step replica 
symmetry breaking result $[S_2^{1\textrm{-}RSB}]/N$ (see section~\ref{rsb-1-section}), 
which is obtained from Eq.~\eqref{renyi}  and Eq.~\eqref{RSB-1-logZ}. Remarkably, 
$[S_2^{1\textrm{-}RSB}]/N$ is in excellent agreement with the extrapolated Monte Carlo 
data for $\beta\lesssim 2$. Finally, for comparison we report in panel (c) the analytical 
result (dotted line) for $S_2/N$ for the infinite-range Ising model without 
disorder at $\omega=1/4$ (see Appendix~\ref{clean-is}).

\section{The classical R\'enyi mutual information}
\label{I2-section}

Here we focus on the behavior of the R\'enyi mutual information $[{\mathcal I}_2]$. 
Similar to $[S_2]$, the mutual information exhibits the volume law $[{\mathcal I}_2]
\propto N$. This is in contrast with local models, where $[{\mathcal I}_n]$, for any 
$n$, by construction obeys an area law at all temperatures. Here we consider the 
mutual information per spin $[{\mathcal I}_2]/N$. 

Figure~\ref{I2-MC} plots $[{\mathcal I}_2]/N$ versus $0\le\beta\le 2.5$ and $\omega=1/2,1/4,
1/8$ (panels from left to right in the Figure). Notice that by definition (cf. Eq.~\eqref{MI}) 
$[{\mathcal I}_n(\omega)]=[{\mathcal I}_n(1-\omega)]$. Circles, squares, and triangles  
are Monte Carlo data for $N=32,64,128$. In the high-temperature region $[{\mathcal I}_2]$ 
exhibits a vanishing behavior. Moreover, finite-size effects are ``small''. Using Eq.~\eqref{logZ-ht} 
it is straightforward to derive the high-temperature behavior of $[{\mathcal I}_n]$ as  
\begin{equation}
[{\mathcal I}_n]=\frac{N\beta^2}{2}\omega(1-\omega)n+{\mathcal O}
(\beta^4).
\end{equation}
$[{\mathcal I}_2]/N$ increases upon lowering the temperature up to $\beta\approx 1$, where 
it exhibits a maximum.  
One should stress that its position is not simply related to the paramagnetic-glassy transition. 
Furthermore, the data for $[{\mathcal I}_2]/N$ at different system sizes do not exhibit any crossing. 
This is in sharp contrast with local models~\cite{jaconis-2013}, where ${\mathcal I}_n/L$ exhibits a 
crossing at a second order phase transition. The dash-dotted line in Fig.~\ref{I2-MC} is the 
analytical result obtained using the replica symmetric (RS) approximation (see 
section~\ref{rs-section}). Formally, this is obtained using Eq.~\eqref{MI} and Eq.~\eqref{rs-F}, 
and it is in perfect agreement with the MC data in the whole paramagnetic phase.  
Finally, we should stress that similar qualitative behavior is observed for ${\mathcal I}_2/N$ 
in the infinite-range Ising model without disorder. The analytical result for ${\mathcal I}_2/N$ 
at $\omega=1/2$ in the thermodynamic limit is reported in panel (a) (dotted line). 

\begin{figure}[t]
\includegraphics*[width=0.93\linewidth]{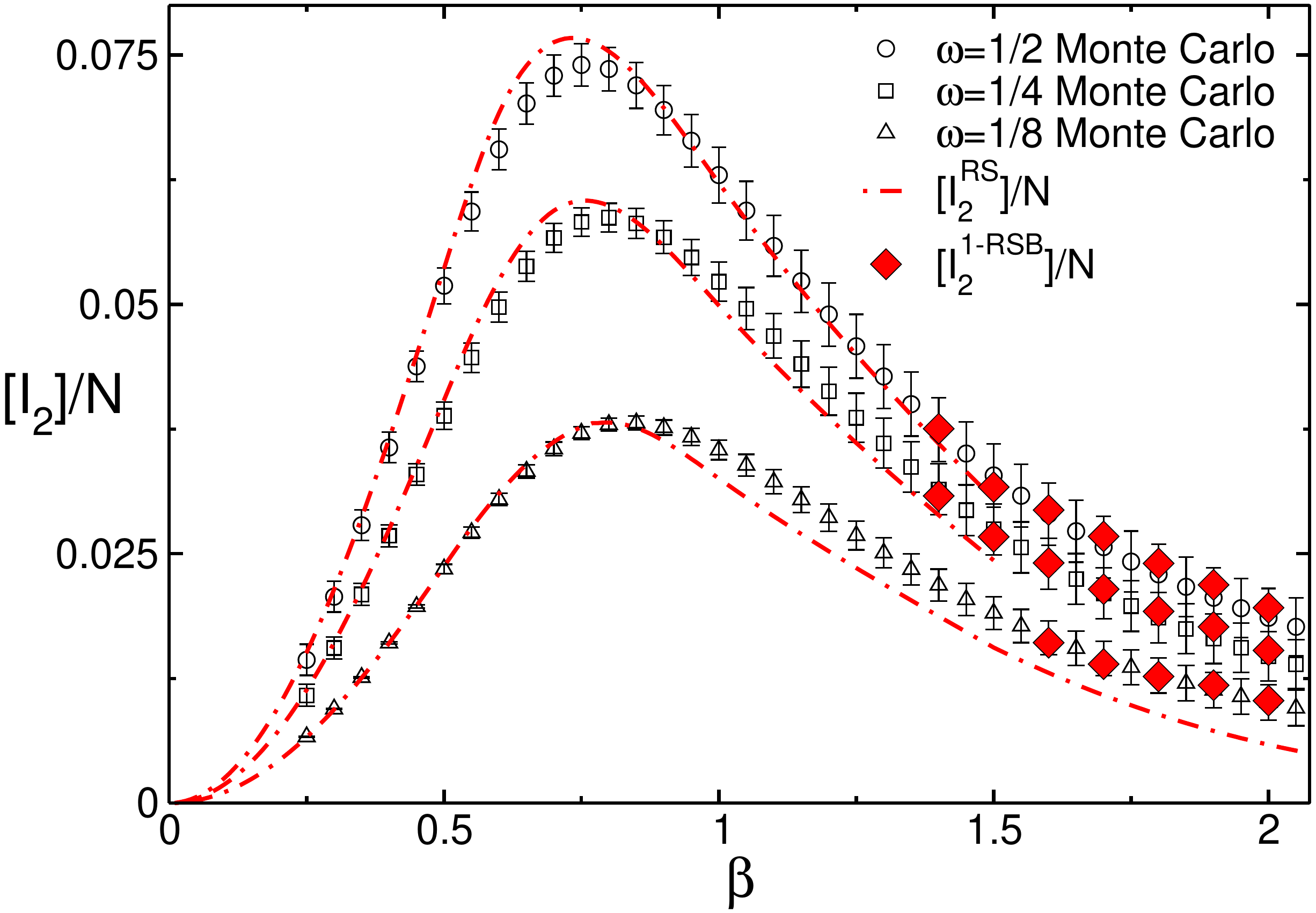}
\caption{The classical mutual information per spin $[{\mathcal I}_2]/N$ 
 in the S-K model on the $2$-sheets booklet: 
 $[{\mathcal I}_2]/N$ plotted versus $\beta$. The symbols denote the 
 Monte Carlo results extrapolated to the thermodynamic limit for 
 booklet ratios $\omega=1/2,1/4,1/8$ (circles, squares, triangles). 
 The dash-dotted lines are the analytical results in the 
 replica-symmetric (RS) approximation. The full symbols (rhombi) 
 are the results in the first-step replica-symmetry-breaking (1-RSB) 
 approximation.
}
\label{I2-extrapolated}
\end{figure}

Interestingly, at low temperatures $[{\mathcal I}_2]/N$ exhibits strong finite-size corrections, 
and significant deviations from the RS result. In order to extract the thermodynamic behavior of 
$[{\mathcal I}_2]/N$ we fit the MC data to 
\begin{equation}
\frac{[{\mathcal I}_2]}{N}=a+\frac{b}{N^{\phi}},
\end{equation}
where we fix $\phi=2/3$. The results of the fits are shown in Fig.~\ref{I2-extrapolated}.  
Different symbols now correspond to different aspect ratios $\omega$. Remarkably, the RS 
approximation (dash-dotted lines) is in good agreement with the extrapolations for $\beta\lesssim 0.6$. 
We should stress that in the region $0.6\lesssim\beta\lesssim 1$, i.e., near the peak, due to the large 
error bars it is difficult to reach a conclusion on the validity of the RS approximation. 
In particular, the data exhibit a systematic shift of the mutual information 
peak towards lower temperatures. Much larger system sizes would be needed to clarify this 
issue. The effect of the replica symmetry breaking, however, 
should be negligible. For instance, at $\beta=0.6$ for $\omega=1/2$ one can estimate that 
$|[{\mathcal I}_2^{RS}]-[{\mathcal I}_2^{1\textrm{-}RSB}]|/N\sim 2\cdot 10^{-5}$. 
Moreover, the results for the infinite-range clean Ising (see Appendix~\ref{is-clean}) suggest 
that logarithmic scaling corrections as $\propto\log(N)/N$ could be present at criticality making 
the extrapolation to the thermodynamic limit tricky. 
Finally, clear deviations 
from the RS result occur at lower temperatures. For instance, for $\omega=1/8$, the numerical results 
exhibit deviations from the RS result already at $\beta\gtrsim 1$. 
These deviations, however, have to be attributed to the physics of the replica symmetry breaking. 
The full rhombi in Fig.~\ref{I2-extrapolated} denote the one-step replica symmetry breaking result 
$[{\mathcal I}_2^{1\textrm{-}RSB}]$, which is obtained using Eq.~\eqref{MI} and Eq.~\eqref{RSB-1-logZ}). 
The agreement between $[{\mathcal I}_2^{1\textrm{-}RSB}]$ and the Monte Carlo data is perfect 
up to $\beta\lesssim 2$.

\section{Summary and Conclusions}
\label{conclusions}

We investigated the \emph{classical} R\'enyi entropy $S_n$ and the mutual information 
${\mathcal I}_n$ in the Sherrington-Kirkpatrick (S-K) model, which is the paradigm model 
of mean-field spin glasses. We focused on the quenched averages $[S_n]$ and $[{\mathcal I}_n]$. 
Specifically, here $[S_n]$ and $[{\mathcal I}_n]$ are obtained from suitable combinations 
of the partition functions of the S-K model on the $n$-sheets booklet (cf. Fig.~\ref{cartoon}). 
This is constructed by ``gluing'' together $n$ independent replicas (``sheets'') of the model. 
On each replica the spins are divided into two groups $A$ and $B$, containing $N_A$ and $N_B$ 
spins respectively. The spins in part $A$ of the different sheets are identified. Due to the 
mean-field nature of the model, physical quantities depend on the bipartition only through the 
aspect ratio $\omega\equiv N_A/N$. 

We first discussed the thermodynamic phase diagram of the S-K model on the $n$-sheets 
booklet, as a function of temperature, and the aspect ratio $\omega$ (cf. Eq.~\eqref{a-ratio}). 
For any fixed $\omega$ the S-K model exhibits a low-temperature glassy phase, which is divided 
by the standard paramagnetic one at high temperatures by a phase transition. 
The critical inverse temperature $\beta_c$ exhibits a non trivial decreasing behavior as a 
function of $\omega$. Moreover, one has $\beta_c=1$ and $\beta_c=1/2$ for $\omega=0$ and 
$\omega=1$, respectively. 
In the high-temperature region the permutation symmetry among both the replicas  and the physical 
sheets  is preserved. This allowed us to provide an exact analytic expression for the free 
energy of the model and several derived quantities, such as the internal energy. 
We compared our results with Monte Carlo simulations, finding perfect agreement. Oppositely, 
in the low-temperature phase the replica symmetry is broken. For instance, we numerically 
observed that the replica-symmetric (RS) result for the internal energy is systematically 
lower than the Monte Carlo data, as in the standard S-K model~\cite{sherrington-1978-prl,
sherrington-1978}. This discrepancy becomes larger upon lowering the temperature.  

Inspired by the Parisi scheme~\cite{parisi-1979}, we devised a systematic way of breaking 
the replica symmetry in successive steps. Our scheme breaks only the symmetry among 
the fictitious replicas, preserving that among the physical ones. Although this appears 
natural, we were not able to provide a rigorous proof that this is the correct symmetry 
breaking pattern. As a consequence, our scheme should be regarded as an approximation, 
and not as an exact solution. Moreover, we restricted ourselves to the one-level replica 
symmetry breaking ($1$-RSB). Surprisingly, the $1$-RSB result for the internal energy 
are in excellent agreement with the Monte Carlo data for  $\beta\lesssim 3$, whereas the 
RS approximation fails already at $\beta\approx 1$. This suggests that the $1$-RSB ansatz 
captures correctly some aspects of the replica symmetry breaking. 

Clear signatures of the replica symmetry breaking can be observed in the behavior of $[S_2]$. 
First, since $[S_2]$ exhibits the volume law behavior $[S_2]\propto N$, we considered its 
density $[S_2]/N$. For finite-size systems, and for any $\omega$, $[S_2]/N$ 
exhibits a maximum at infinite temperature, and it is a decreasing function of the temperature, 
as expected. Finite-size corrections are negligible at high temperatures, whereas they increase 
upon lowering the temperature. In the paramagnetic phase we were able to determine the functional 
form of $[S_2]/N$ in the thermodynamic limit, using the replica-symmetric approximation. This 
perfectly matched the Monte Carlo data. At low temperatures deviations from the RS result are 
present, reflecting the replica symmetry breaking. Remarkably, the one-step replica symmetry 
breaking ($1$-RSB) result $[S_2^{1\textrm{-}RSB}]/N$ fully describes the Monte Carlo 
data for $\beta\lesssim 3$, consistent with what was observed for the internal energy. 

Finally, we considered the R\'enyi mutual information $[{\mathcal I}_2]$. This obeys a volume 
law for any $\beta$ and $\omega$, in contrast with local spin models, where an area law is 
observed~\cite{wolf-2008}. The corresponding density $[{\mathcal I}_2]/N$ vanishes in 
both the infinite-temperature and the zero-temperature limits. Surprisingly, $[{\mathcal I}_2]/N$ 
does not exhibit any crossing for different system sizes at the paramagnetic-glassy transition, 
in striking contrast with local spin models~\cite{jaconis-2013}. For any $\omega$, $[{\mathcal I}_2]/N$ 
exhibits a maximum for $\beta\approx 1$. The position of this maximum is not simply 
related to the paramagnetic-glassy transition. At high temperature $[{\mathcal I}_2]/N$ 
is described analytically by the RS result $[{\mathcal I}_2]^{RS}/N$. Deviations from the RS result, 
if present, are $\lesssim 10^{-3}$. On the other hand, at low temperatures 
one has to include the effects of the replica symmetry breaking. Similar to $[S_2]$, the $1$-RSB 
approximation $[{\mathcal I}^{1\textrm{-}RSB}_2]/N$ is in good agreement with the Monte Carlo 
data for $\beta\lesssim 3$.

Our work opens several research directions. First, it would be interesting to extend our 
results taking into account the full breaking of the replica symmetry, i.e., going beyond 
the one-step replica symmetry breaking approximation. This would allow us to reach a 
conclusion on the correctness of the replica symmetry breaking scheme that we used. Moreover, 
it would be interesting to discuss the finite size-corrections to the saddle point 
approximation. An intriguing direction would be to investigate whether the glassy critical 
behavior  is reflected in the volume-law corrections to the classical R\'enyi entropies 
and mutual information. Finally, it would be interesting to extend our results to {\it quantum} 
spin systems exhibiting glassy behavior and replica symmetry breaking~\cite{read-1995,
andreanov-2012}.

\section{Acknowledgements}
We would like to thank Pasquale Calabrese for useful discussions. V.A.  acknowledges  
financial support from the ERC under Starting Grant 279391 EDEQS. S.I. and L.P. 
acknowledge support  from the  FP7/ERC Starting Grant No. 306897.

\appendix

\section{The saddle point equations} 
\label{saddle-equations}

In this section we provide the analytical expression for the saddle point equations 
\eqref{saddle-final}\eqref{saddle-final-a}, which determine the overlap tensor 
$q_{\gamma\gamma'}^{rr'}$ (see section~\ref{replica-sec}). We restrict ourselves 
to zero magnetic field and to the $2$-sheets booklet (see Fig.~\ref{cartoon}). It is 
straightforward to generalize the calculation to the case with non zero magnetic 
field and to the $n$-sheets booklet. Here we provide the saddle point equations 
for both the replica-symmetric (RS) (see section~\ref{rs-section}) and the one-step 
replica symmetry breaking ($1$-RSB) approximations (see section~\ref{rsb-1-section}). 

\subsection{The replica-symmetric (RS) approximation}

In the replica-symmetric approximation $q_{\gamma\gamma'}^{rr'}$ depends on the two 
parameters $q_0,q_0'\in\mathbb{R}$ (cf. Eq.~\eqref{rs-ansatz}). The saddle point 
equations are derived from the RS approximation for the free energy $[F_{RS}
(\omega,n,\beta)]$ (cf. Eq.~\eqref{rs-F}) as $\nabla_{\mathbf{q}}[F_{RS}(\omega,2,
\beta)]=0$, where $\mathbf{q}\equiv(q_0,q_0')$, and $\nabla_{\mathbf{q}}\equiv
\partial/\partial\mathbf{q}$. A straightforward calculation gives 
\begin{equation}
\label{RS-saddle-1}
\frac{1-q_0}{1-\omega}=2\int dz G_0(z)(1+\exp(2\beta^2(q_0-q_0'))\cosh(2z))^{-1},
\end{equation}
and
\begin{multline}
\label{RS-saddle-2}
2\beta^2\frac{1-q_0'}{1-\omega}=
\int dz G_0(z)\Big\{\frac{\omega}{1-\omega}\Delta_0(z)
\log\cosh(2z)\\
+\Delta_0(z)\log(1+\exp(2\beta^2(q_0-q_0'))\cosh(2z))\\
+2\beta^2(1+\exp(2\beta^2(q_0-q_0'))\cosh(2z))^{-1}
\Big\},
\end{multline}
where we defined $\Delta_0(z)$ as 
\begin{equation}
\Delta_0(z)\equiv\frac{z^2}{2\beta^2q_0'^2}-\frac{1}{2q_0'},
\end{equation}
and the so-called heat kernel $G_0(z)$ as 
\begin{equation}
G_0(z)\equiv\frac{1}{\sqrt{2\pi \beta^2 q_0'}}\exp\Big(-\frac{z^2}
{2\beta^2 q_0'}\Big).
\end{equation}
Notice that $\int dz G_0(z)\Delta_0(z)=0$.

\subsection{The one-step replica symmetry breaking ($1$-RSB) approximation}

In the one-step replica symmetry breaking approximation (see section~\ref{rsb-1-section}) 
$q_{\gamma\gamma'}^{rr'}$ depends on the four parameters $q_0,q_0',q_1',m_1\in\mathbb{R}$. 
The saddle point equations are given as $\nabla_{\mathbf{p}}[F_{1\textrm{-}RSB}]=0$, 
where $\mathbf{p}\equiv(q_0,q_0',q_1',m_1)$, and $[F_{1\textrm{-}RSB}]$ is 
the disorder-averaged free energy given in Eq.~\eqref{RSB-1-logZ}.
It is useful to define the modified heat kernel $G_1(z)$ as 
\begin{equation}
G_1(z)\equiv\frac{1}{\sqrt{2\pi \beta^2 (q_1'-q_0')}}\exp\Big(-\frac{z^2}
{2\beta^2 (q_1'-q_0')}\Big), 
\end{equation}
and 
\begin{align}
& \Delta_1(z)\equiv \frac{z^2}{\beta^2(q_1'-q_0')^2}-\frac{1}{q_1'-q_0'}\\\nonumber
& \Gamma(z)\equiv \Big\{\int dz'G_1(z')\cosh^{m_1}(2z+2z')\Big\}^{-1}\\\nonumber
& \Gamma'(z)\equiv \Big\{\int dz'G_1(z')\Big(1+\cosh(2z+2z')\Big)^{m_1}\Big\}^{-1}\\\nonumber
& \Theta(z,z')\equiv 1+\exp(2\beta^2(q_0-q_1'))\cosh(2z+2z'). 
\end{align}
Finally, the saddle point equations for $q_0,q_0',q_1',m_1$ are obtained as 
\begin{widetext}
\begin{multline}
0=(1+q_0-2\omega)\exp(-2\beta^2(q_0-q_1'))
-2(1-\omega)\int dzdz'G_{0}(z)
\Gamma'(z)G_{1}
(z')\cosh(2z+2z')\Theta^{m_1-1}(z,z')
\label{RSB-1-saddle-a}
\end{multline}
\begin{multline}
0=4\beta^2m_1q_0'+\frac{\omega}{m_1}\int dz G_{0}(z)
\Delta_0(z)\log\int dz'G_{1}(z')
\cosh^{m_1}(2z +2z')\\
+\frac{1-\omega}{m_1}\int dzG_{0}(z)
\Delta(z)\log\int dz'G_{1}(z')\Theta^{m_1}(z,z')
-\frac{\omega}{m_1}\int dzdz'G_{0}(z)\Gamma(z)
G_{1}(z')\Delta_1(z')\cosh^{m_1}(2z+2z')\\
-\frac{1-\omega}{m_1}
\int dzdz'G_{0}(z)\Gamma'(z)G_{1}(z')\Delta_1(z')\Theta^{m_1}(z,z')
\label{RSB-1-saddle-b}
\end{multline}
\begin{multline}
0=-4\beta^2(m_1-1)q_1'-4\beta^2\omega+\frac{\omega}{m_1}
\int dzdz'G_{0}(z)\Gamma(z)G_{1}(z')\Delta_1(z')
\cosh^{m_1}(2z+2z')\\+
\frac{1-\omega}{m_1}\int dzdz'G_{0}(z)
\Gamma'(z)
G_{1}(z')\Big\{\Delta_1(z')\Theta^{m_1}(z,z')
-4\beta^2m_1\exp(2\beta^2(q_0-q_1'))\cosh(2z+2z')
\Theta^{m_1-1}(z,z')
\Big\}
\label{RSB-1-saddle-c}
\end{multline}
\begin{multline}
0=-\beta^2(q_1'^2-q_0'^2)+\frac{\omega}{m_1}
\int dzdz'G_{0}(z)\Gamma(z)G_{1}(z')\cosh^{m_1}(2z+2z')
\log\cosh(2z+2z')\\+
\frac{1-\omega}{m_1}\int dzdz' G_{0}(z)\Gamma'(z)
G_{1}(z')\Theta^{m_1}(z,z')\log(\Theta(z,z'))
-\frac{1-\omega}{m_1^2}\int dz G_{0}(z)\log\int dz' 
G_{1}(z')\Theta^{m_1}(z,z')\\
-\frac{\omega}{m_1^2}\int dzG_{0}(z)\log
\int dz'G_{1}(z')\cosh^{m_1}(2z+2z').
\label{RSB-1-saddle-d}
\end{multline}
\end{widetext}
%

\section{Mutual information in the infinite-range clean Ising model}
\label{clean-is}

Here we discuss the R\'enyi entropy $S_2$ and the associated mutual information 
${\mathcal I}_2$ in the infinite-range Ising model without disorder, defined by 
the hamiltonian 
\begin{equation}
\label{is-clean}
{\mathcal H}=-\frac{J}{N}\sum\limits_{i<j}S_iS_j+h\sum_i S_i.
\end{equation}
We consider the situation without external magnetic field, i.e., $h=0$ and 
fix $J=1$. The phase diagram of the model exhibits a paramagnetic phase at 
high temperature, while at low temperatures there is ferromagnetic order. The 
two phases are separated by a phase transition at $\beta_c=1$.
We consider both the situations with finite $N$ as well as the 
thermodynamic limit. For finite $N$ we provide exact numerical results for 
$S_2$ and ${\mathcal I}_2$, while we address the thermodynamic limit using 
the booklet construction (see~\ref{booklet}) and the saddle point 
approximation. 

Similar to the S-K model, we find that the para-ferro transition persists on 
the booklet, and the critical temperature depends on the booklet ratio 
$\omega$. The R\'enyi entropy $S_2$ and the mutual information ${\mathcal 
I}_2$ are extensive, and their densities $S_2/N,{\mathcal I}_2/N$ finite 
and smooth at any temperature. Notice that this different for the Shannon 
mutual information~\cite{wilms-2012} ${\mathcal I}_1$, which is {\it finite} 
(i.e., ${\mathcal I}_1\sim {\mathcal O}(1)$) in the whole phase diagram, 
except for a logarithmically divergent behavior (with system size) at the 
critical point $\beta_c=1$. 

Finally, we show that the information about the phase transition is encoded in 
the subleading, i.e. $o(N)$, corrections of ${\mathcal I}_2$. Precisely, 
logarithmically divergent contributions are present at $\beta=1/2,1$ and 
$\beta=\beta_c(\omega)$, with $\beta_c(\omega)$ the critical temperature of the 
model on the booklet at fixed aspect ratio $\omega$.

\subsection{Saddle point approximation}

We first discuss the thermodynamic limit. The calculation of the booklet partition 
function $Z(\omega,n,\beta)$ (see section~\ref{booklet}) can be done using the same 
techniques as in section~\ref{the-model}. The result reads 
\begin{multline}
\label{clean-is-Z}
Z(\omega,n,\beta)=
\exp\Big(-\frac{\beta}{2}n\Big)\Big(2\pi\frac{\beta}{N}\Big)^{-\frac{n}{2}}
\int\prod_{\delta}dz_\delta\exp N\Big\{\\
-\sum_\delta\Big[\frac{z^2_\delta}{2\beta}
-(1-\omega)\log(2\cosh z_\delta)\Big]+\omega\log(2\cosh\sum_\delta z_\delta)
\Big\}, 
\end{multline}
with $\delta=1,\dots,n$. Using the saddle point approximation, from~\eqref{clean-is-Z} 
one obtains $F(\omega,n,\beta)\equiv\log Z(\omega,n,\beta)$ as 
\begin{multline}
\label{clean-is-F}
F(\omega,n,\beta)\approx
N\Big\{-\sum_\delta\Big[\frac{z^2_\delta}{2\beta}
-(1-\omega)\log(2\cosh z_\delta)\Big]\\
+\omega\log(2\cosh\sum_\delta z_\delta)\Big\}, 
\end{multline}
where we neglected subleading contributions $o(N)$. The parameters 
$\{z_\delta\}_{\delta=1}^n$ are solutions of the saddle point equations 
\begin{equation}
\label{clean-is-spe}
0=z_\delta-\beta(1-\omega)\tanh z_\delta-\beta\omega\tanh\big(\sum_\delta 
z_\delta\big),\quad \forall\delta.
\end{equation}
For $n=2$, the system~\eqref{clean-is-spe} becomes 
\begin{align}
& 0=z_1-\beta(1-\omega)\tanh z_1-\beta\omega\tanh\big(z_1+z_2\big)\\
& 0=z_2-\beta(1-\omega)\tanh z_2-\beta\omega\tanh\big(z_1+z_2\big). 
\end{align}
The system has the solutions $z_1=z_2=z_s=0$ for $\beta\le\beta_c=1/(1+\omega)$, 
and $z_1=z_2=z_s\ne 0$ for $\beta>\beta_c$. Note that $\beta_c$ is different 
in presence of disorder (see~\eqref{tc}). 

\begin{figure}[t]
\includegraphics*[width=0.9\linewidth]{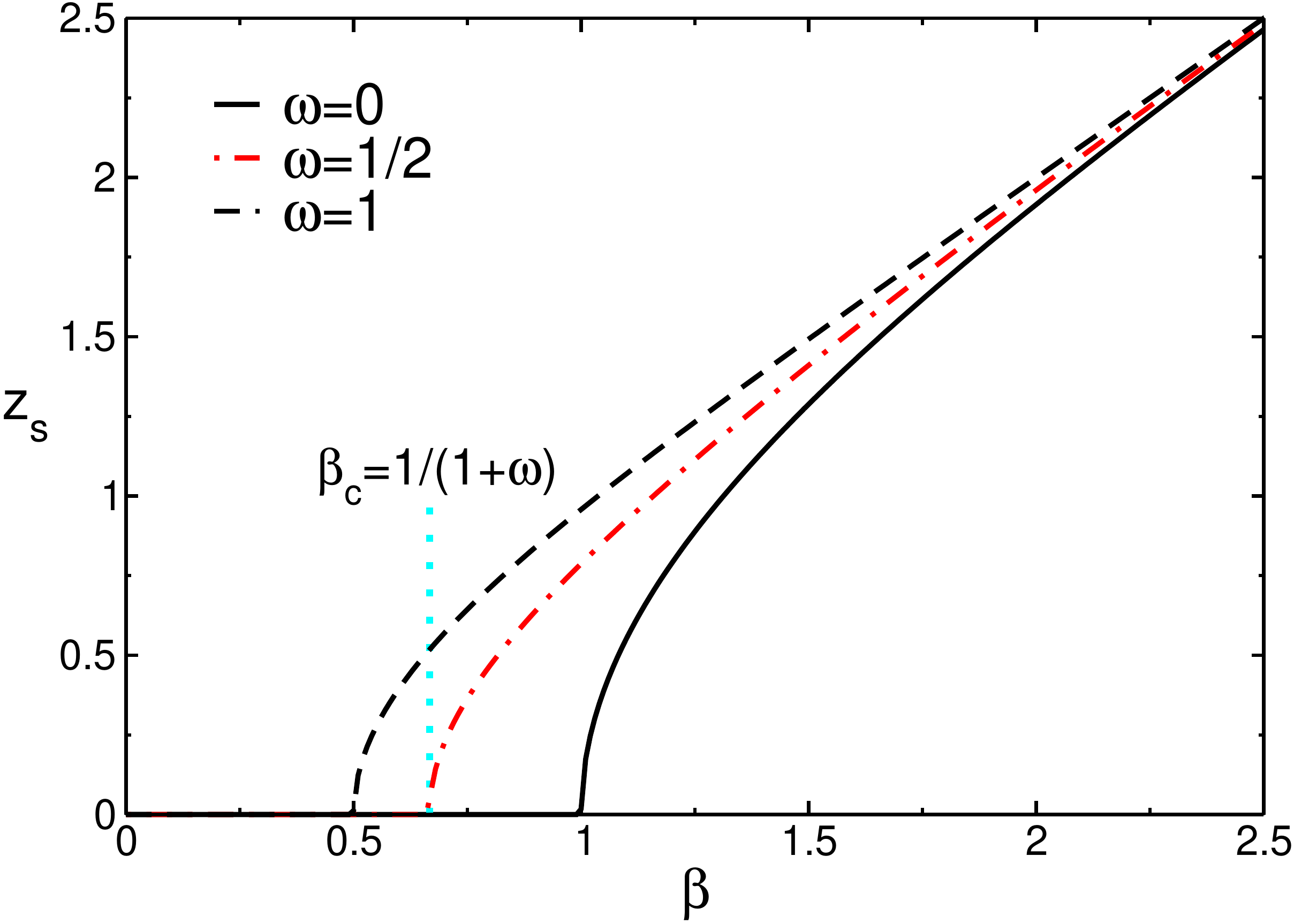}
\caption{ The infinite-range Ising model on the $2$-sheets booklet: saddle point 
 approximation. The solution $z_s$ of the saddle point equations~\eqref{clean-is-spe} 
 plotted as a function of the inverse temperature $\beta$ and for booklet aspect 
 ratio $\omega=0,1/2,1$. The vertical dotted line denotes the critical inverse 
 temperature for $\omega=1/2$. 
}
\label{saddle_clean}
\end{figure}

The solution $z_s$ of the saddle point equations~\eqref{clean-is-spe} is shown 
in Fig.~\ref{saddle_clean} as a function of $\beta$ for $\omega=0,1/2,1$. 
The vertical-dotted line is $\beta_c$ for $\omega=1/2$. Notice that at low 
temperatures $z_s$ is independent on $\omega$, and one has $z_s\sim\beta$. 

\subsection{R\'enyi entropy and mutual information: saddle point results}

\begin{figure}[t]
\includegraphics*[width=0.9\linewidth]{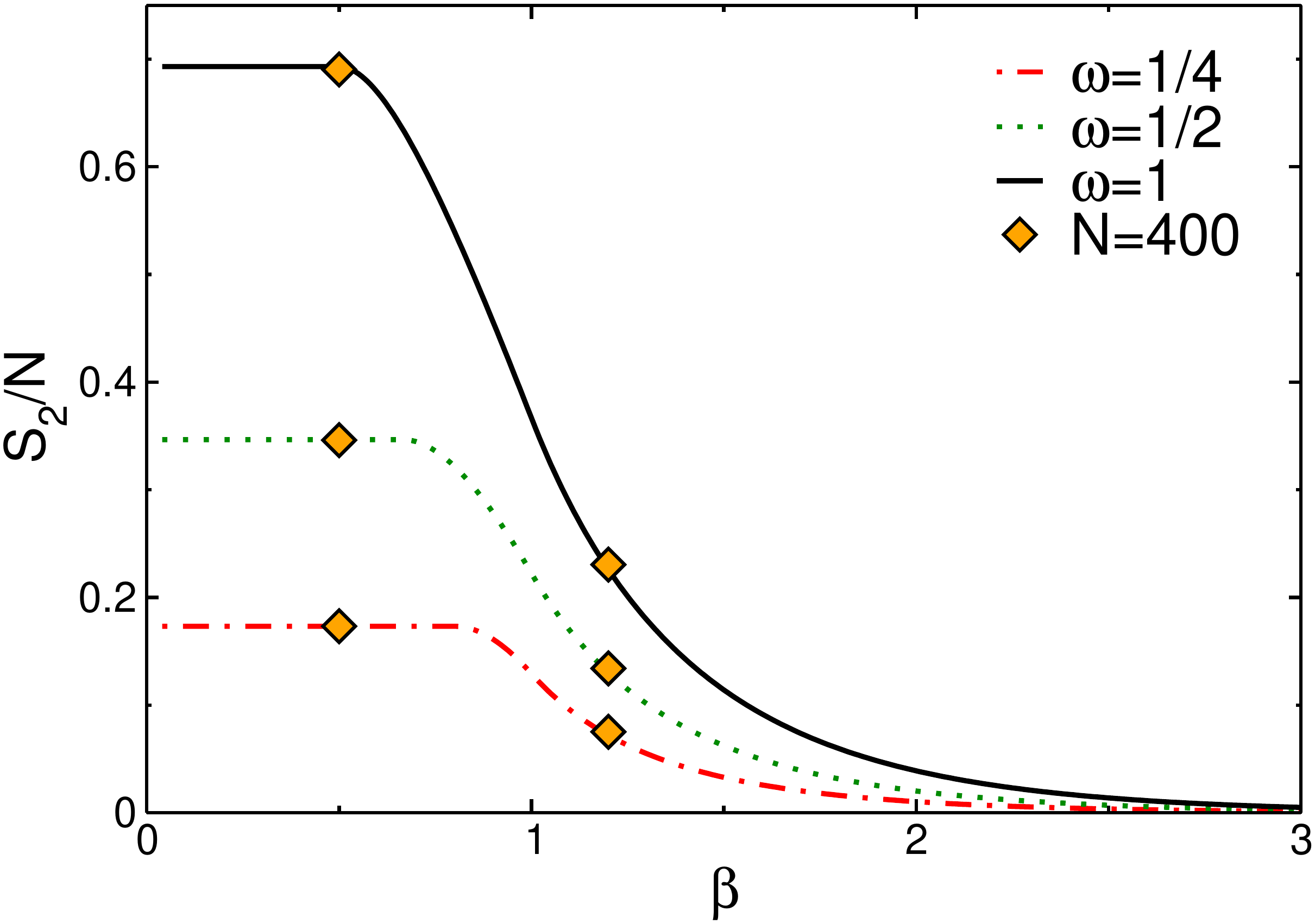}
\caption{ The infinite-range Ising model on the $2$-sheets booklet in the thermodynamic 
 limit: the classical R\'enyi entropy density $S_2/N$, with $N$ the total number of spins 
 on a single sheet of the booklet. $S_2/N$ is plotted against the inverse temperature 
 $\beta$ and for booklet aspect ratio $\omega=1/4,1/2,1$. Note tht at high temperature 
 $S_2=\omega\log(2)$. The lines are obtained using the saddle point approximation, while 
 the full symbols denote the exact numerical results for a finite system with $N=400$. 
}
\label{renyi_clean}
\end{figure}

It is straightforward to obtain the R\'enyi entropies $S_n=(F(\omega=0,n,\beta)-
F(\omega,n,\beta))/(n-1)$ using~\eqref{clean-is-F} and~\eqref{renyi}. 
Figure~\ref{renyi_clean} plots the entropy density $S_2/N$ versus the inverse 
temperature $\beta$. $S_2$ exhibits a volume-law behavior $S_2\propto N$ for 
all values of $\beta$ and $\omega$. The different curves correspond to different 
booklet aspect ratios $\omega$. For $\beta\le\beta_c$ one has the flat behavior 
$S_2=\omega\log(2)$, whereas $S_2$ is vanishing in the low-temperature regime. 
The full symbol (rhombi) in Figure~\ref{renyi_clean} are exact results for a 
booklet with $N=400$ spins per sheet (see next section), and are in good 
agreement with the saddle point approximation results.  

\begin{figure}[t]
\includegraphics*[width=0.9\linewidth]{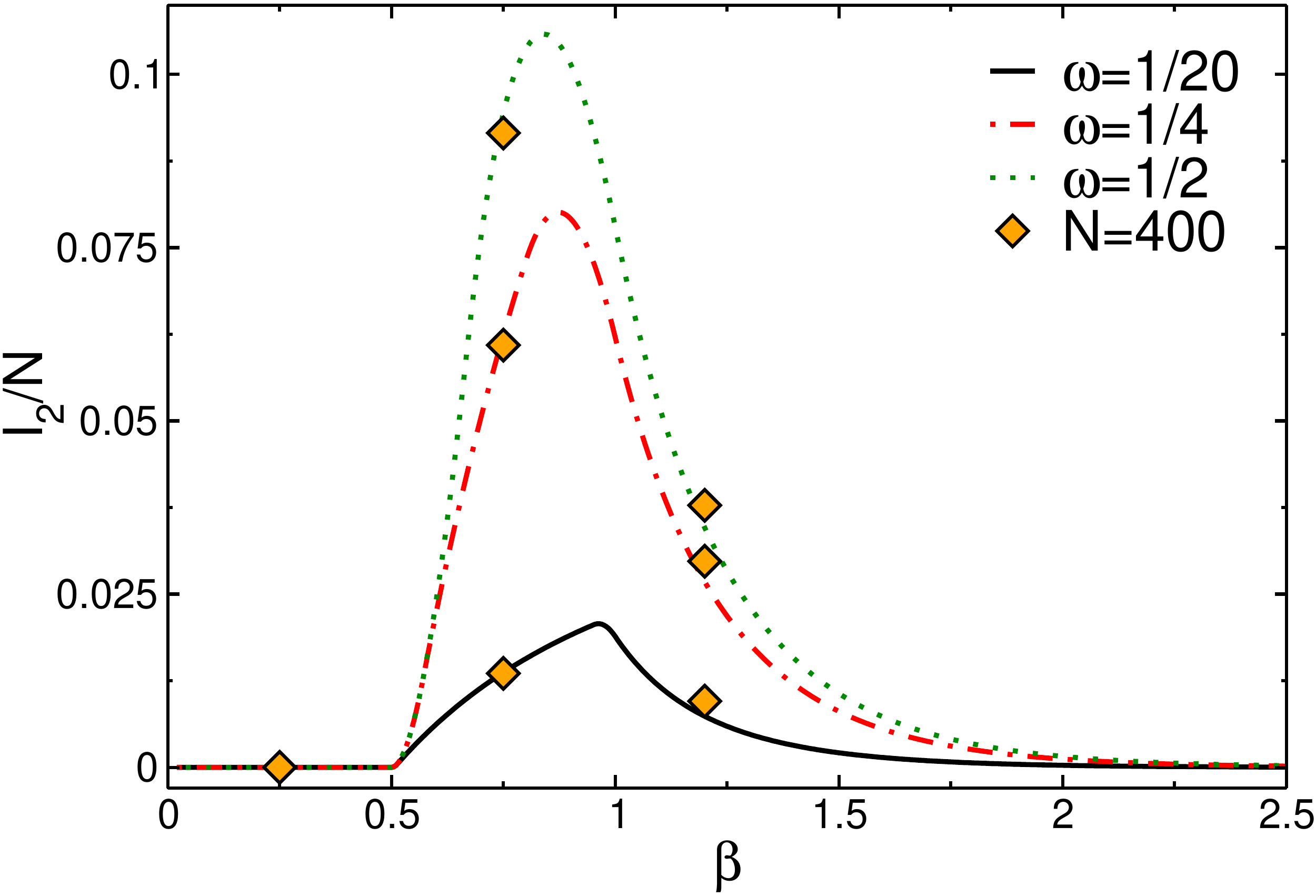}
\caption{ The infinite-range Ising model on the $2$-sheets booklet in the thermodynamic 
 limit: the classical mutual information density ${\mathcal I}_2/N$ plotted against the 
 inverse temperature $\beta$ and for booklet aspect ratio $\omega=1/4,1/2,1$. Note that  
 ${\mathcal I}_2=0$ for $\beta\le 1/2$. The full symbols are the the exact numerical 
 results for a finite system with $N=400$. 
}
\label{mi_clean}
\end{figure}

The mutual information per spin ${\mathcal I}_2/N$ is reported in 
Figure~\ref{mi_clean}. The lines are the analytical results for booklet aspect 
ratios $\omega=1/20,1/4,1/2$, while the full symbols are the exact results for 
the booklet with $N=400$ spins per sheet. Clearly, ${\mathcal I}_2/N$ is exactly zero 
in the high-temperature region for $\beta\le1/2$, it exhibits at maximum 
at $\beta\approx 1$, and it is vanishing in the limit $\beta\to\infty$. 

Finally, one should stress that a dramatically different behavior is observed 
in the Shannon mutual information ${\mathcal I}_1\equiv \lim_{n\to 1}{\mathcal I}_n$. 
Specifically, ${\mathcal I}_1$ is finite in the thermodynamic limit, and it only 
exhibits a logarithmic divergence as ${\mathcal I}=1/4\log(N)$ at the 
critical point $\beta_c=1$ (see Ref.~\onlinecite{wilms-2012}).

\subsection{R\'enyi entropy and mutual information: Exact treatment}

For finite $N$, $S_2$ and ${\mathcal I}_2$ can be calculated exactly using the 
results obtained in Ref.~\onlinecite{wilms-2012}. The eigenstates 
of~\eqref{is-clean} are product states. They can be characterized as $|p,i\rangle$, 
with $p$ the total number of up spins (the remaining $N-p$ being down spins) and $i$ 
labelling the $B(N,p)\equiv N!/(p!(N-p)!)$ eigenstates with the same $p$. 
The partition function of~\eqref{is-clean} is given as 
\begin{equation}
Z(\beta)\equiv\textrm{Tr}\rho=\sum\limits_{p=0}^NB(N,p)e^{2N\beta(\frac{p}{N}-
\frac{1}{2})^2}.
\end{equation}
Similarly, one has 
\begin{equation}
\label{rhon}
Z^n(\beta)\equiv\textrm{Tr}\sum\limits_{p=0}^NB(N,p)e^{2nN\beta(\frac{p}{N}-
\frac{1}{2})^2}. 
\end{equation}
The thermal density matrix $\rho$ is defined as $\rho\equiv 1/Ze^{-\beta{\mathcal H}}$. 
Given a bipartition of the spins into two groups $A$ and $B$ containing $N_A$ 
and $N-N_A$ spins, respectively, the reduced density matrix for $A$ is obtained 
as~\cite{wilms-2012} 
\begin{equation}
\rho_A\equiv\frac{1}{Z}\textrm{Tr}_B\sum\limits_{p,i}e^{2N\beta(\frac{p}{N}-
\frac{1}{2})^2}|p,i\rangle\langle p,i|. 
\end{equation}
By performing the trace over part $B$ one obtains 
\begin{equation}
\label{rhoA}
\rho_A=\sum\limits_{p_A,i_A}R(p_A)|p_A,i_A\rangle\langle p_A,i_A|,
\end{equation}
where $|p_A,i_A\rangle$ form a basis for part $A$ of the system and 
\begin{equation}
R(p_A)\equiv\frac{1}{Z}\sum\limits_{p_B=0}^{N-N_A}B(N-N_A,p_B)e^{2N\beta
(\frac{p_A}{N}+\frac{p_B}{N}-\frac{1}{2})^2}.
\end{equation}
From~\eqref{rhoA} it is straightforward to obtain 
\begin{equation}
\label{trrhoA}
\textrm{Tr}\rho_A^n=\sum\limits_{p_A=0}^{N_A}B(N_A,p_A)R^n(p_A). 
\end{equation}
Similarly, $\textrm{Tr}\rho^n_B$ is obtained from~\eqref{trrhoA} substituting 
$A\to B$ and $N_A\to N-N_A$, while $\textrm{Tr}\rho^n_{A\cup B}\equiv\textrm{Tr}
\rho^n$. The R\'enyi entropies $S_n$ and the mutual 
informations ${\mathcal I}_n$ can be calculated numerically 
using~\eqref{trrhoA}, \eqref{rhon} and the definitions~\eqref{renyi-intro}
\eqref{MI}. The numerical results for $S_2$ and ${\mathcal I}_2$ for a system with 
$N=400$ spins are shown in Figure~\ref{renyi_clean} and Figure~\ref{mi_clean}, and 
are in good agreement with the results in the thermodynamic limit. 

\begin{figure}[t]
\includegraphics*[width=0.9\linewidth]{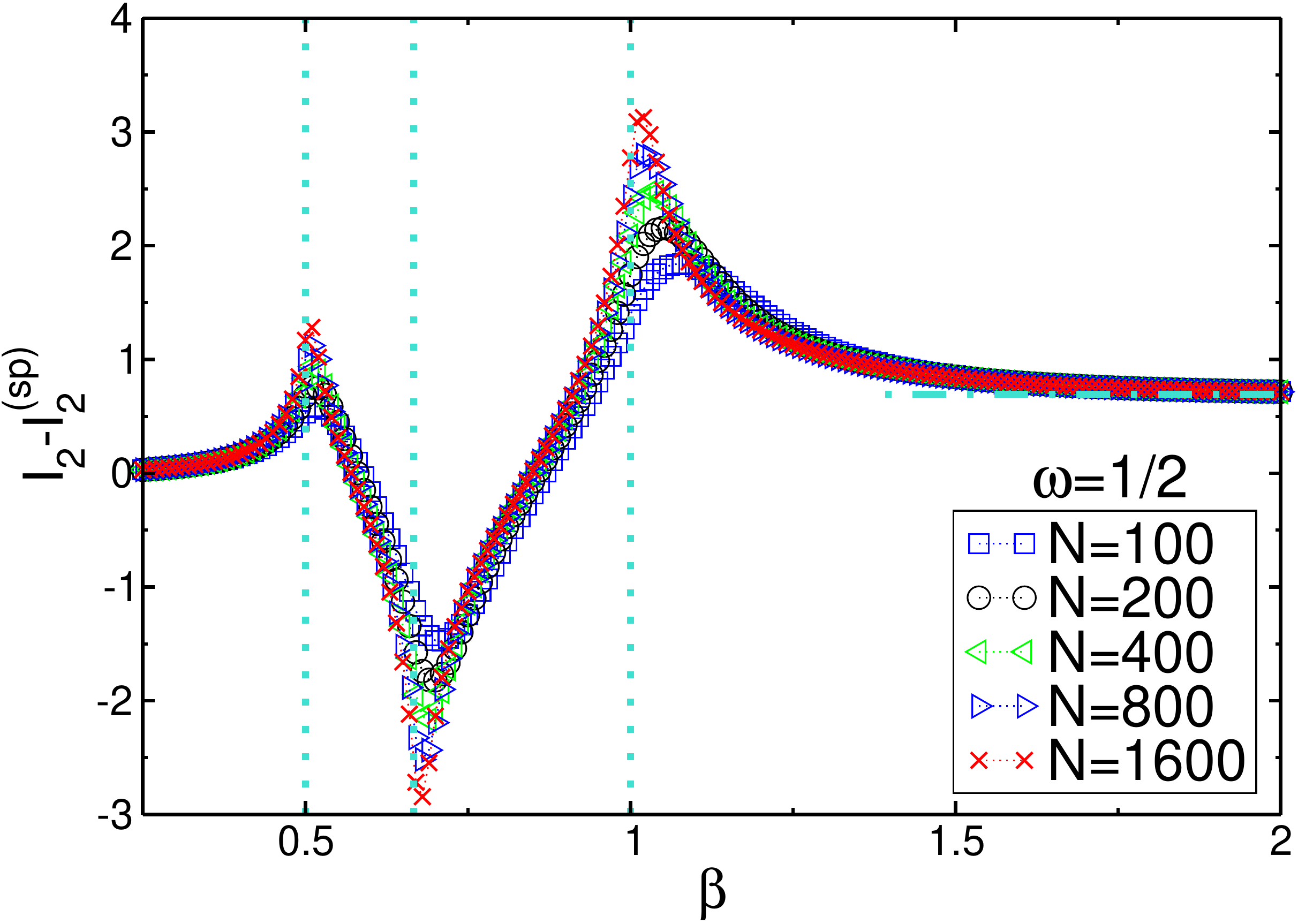}
\caption{ The infinite-range Ising model on the $2$-sheets booklet: saddle point corrections. 
 The correction for the mutual information ${\mathcal I}_2-{\mathcal I}_2^{(sp)}$ are 
 plotted against the inverse temperature $\beta$. The data are for a booklet with aspect 
 ratio $\omega=1/2$. Here ${\mathcal I}_2^{(sp)}$ are the same data as in Figure~\ref{mi_clean} 
 (dotted line). The vertical lines denote the critical temperatures $\beta_c$ for $\omega=1,
 1/2,0$ (from left to right). The horizontal dash-dotted line is $\log(2)$. Note the 
 (logarithmically) divergent behavior at $\beta_c$. 
}
\label{sp_corr}
\end{figure}

\begin{figure}[t]
\includegraphics*[width=0.9\linewidth]{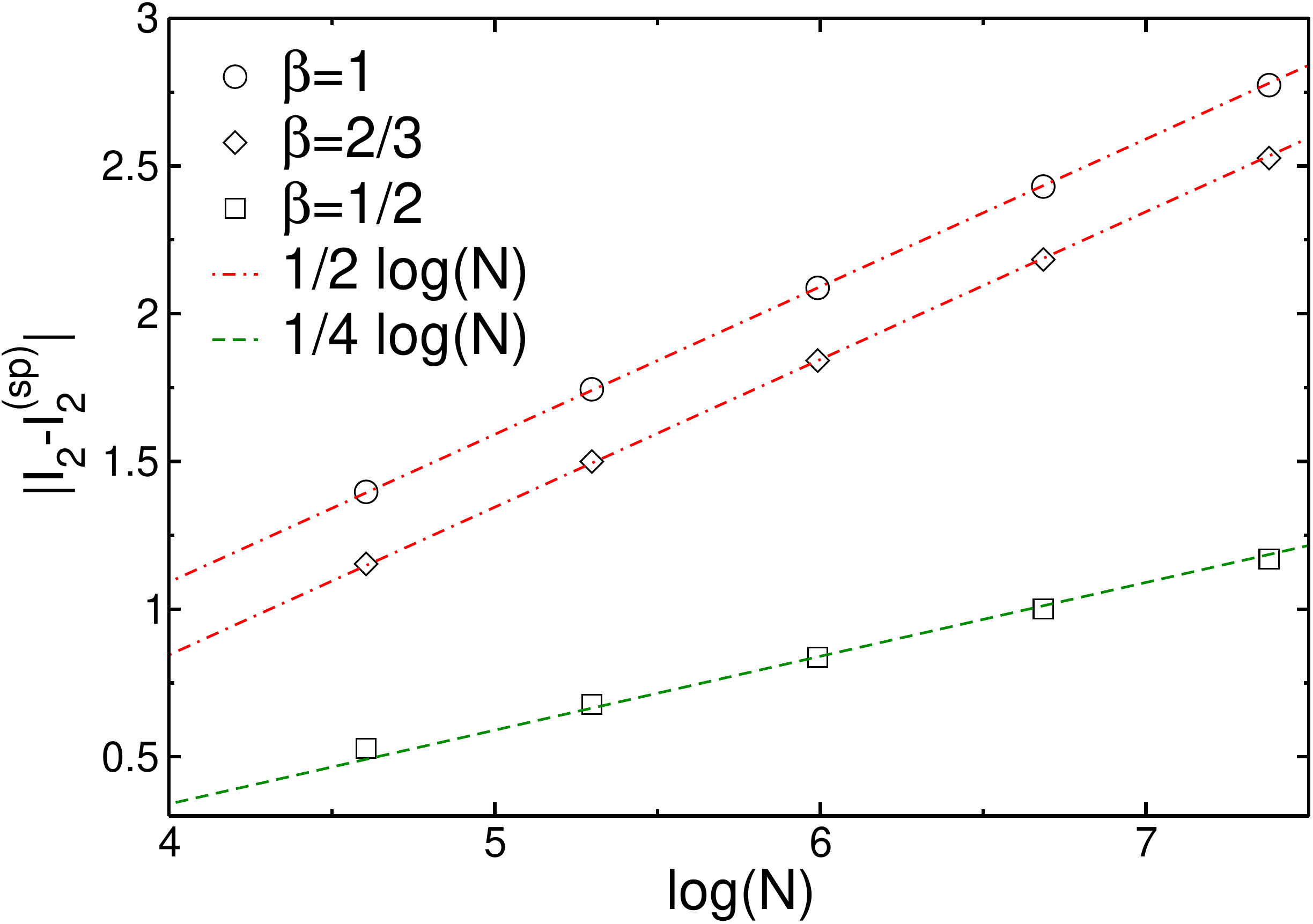}
\caption{ The infinite-range Ising model on the $2$-sheets booklet: 
 Scaling of the corrections of the mutual information ${\mathcal I}_2$ 
 at the booklet critical points $\beta_c=1/2,2/3,1$ (see 
 vertical lines in Figure~\ref{sp_corr}). We plot $|{\mathcal I}_2-
 {\mathcal I}_2^{(sp)}|$ versus $\log(N)$, with $N$ the number of spins 
 on one sheet of the booklet. Here ${\mathcal I}_2^{(sp)}$ is the same 
 as in Figure~\ref{mi_clean} (dotted line).  
}
\label{clean_log}
\end{figure}

Figure~\ref{sp_corr} focuses on the finite-size subextensive corrections for 
${\mathcal I}_2$. We plot ${\mathcal I}_2-{\mathcal I}^{(sp)}_2$, with ${\mathcal I}_2^{(sp)}$ 
denoting the saddle point extensive part of ${\mathcal I}_2$. We restrict ourselves to 
$\omega=N_A/N=1/2$, although similar results have to be expected at different $\omega$. 
Clearly, ${\mathcal I}_2-{\mathcal I}_2^{(sp)}$ is vanishing at high temperatures, 
whereas one has ${\mathcal I}_2-{\mathcal I}_2^{(sp)}\to\log(2)$ in the limit 
$\beta\to\infty$ (horizontal line). Surprisingly, ${\mathcal I}_2-{\mathcal I}_2^{(sp)}$ 
diverges in the thermodynamic limit for $\beta=1/2,2/3,1$ (vertical lines in the 
Figure), which are the critical temperatures for the model on a booklet with 
$\omega=1,1/2,0$. 
This divergence is logarithmic as a function of $N$, as confirmed in Figure~\ref{clean_log} 
plotting $|{\mathcal I}_2-{\mathcal I}_2^{(sp)}|$ versus $\log(N)$ for fixed $\beta=1/2,2/3,1$. 
Interestingly, the precise behavior depends on $\beta$. One has 
$|{\mathcal I}_2-{\mathcal I}_2^{(sp)}|\propto 1/4\log(N)$ for $\beta=1/2$ and 
$|{\mathcal I}_2-{\mathcal I}_2^{(sp)}|\propto 1/2\log(N)$ for $\beta=2/3$ and 
$\beta=1$. Again, this is dramatically different for the von Neumann mutual 
information ${\mathcal I}_1=\lim_{n\to 1}{\mathcal I}_n$, which exhibits only one 
divergent peak~\cite{wilms-2012} at $\beta=1$.

\section{Monte Carlo method to calculate the classical R\'enyi entropies}
\label{mc-method}

Here we describe the Monte Carlo method that we used to calculate the R\'enyi entropies 
$S_n$ and the mutual information ${\mathcal I}_n$. The method exploits the representation 
of $S_n$ as ratio of partition functions as 
\begin{equation}
\label{renyi-app}
S_n(A)\equiv \frac{1}{1-n}\log\left(\frac{Z(A,n,\beta)}{Z^n(\beta)}\right),
\end{equation}
where $Z(A,n,\beta)$ and $Z(\beta)$ are the partition functions of the model on the booklet and 
on the plane, respectively (see Fig.~\ref{cartoon} and section~\ref{booklet} for the definitions).  
There are only few approaches to numerically calculate the ratio of partition functions in 
Eq.~\eqref{renyi-app}. For instance, a brute force numerical integration of the internal energy as 
a function of temperature\cite{jaconis-2013} can be used to calculate $Z(A,n,\beta)$ and $Z(\beta)$. 
However, this requires high accuracy over a large range of temperature.

Here we directly measure the ratio of partition functions using the so-called \emph{ratio trick}. In 
the ratio trick one splits the subsystem $A$ in a set of subsystems $A_i$
such that $A_i\subset A_{i+1}$, and the largest subsystem in the set is simply $A$ itself. Then one writes 
\begin{align}
\label{ratio-trick}
\frac{Z(A,n,\beta)}{ Z^n(\beta)} = \prod_{i=0}^{N/4-1} \frac{Z(A_i,n,\beta)}{Z(A_{i+1},n,\beta)},
\end{align}
where for simplicity we specialized to intervals $A_i$ such that
\begin{align}
\label{sub-choice}
A_i = 4i,\quad i \in [0,N/4].
\end{align}
Crucially, if the length of $A_i$ increases mildly with $i$, each term in the product 
in the right-hand side of Eq.~\eqref{ratio-trick} can be sampled efficiently in Monte 
Carlo~\cite{stephan-2014}. Notice that the same trick has also been used in 
Ref.~\onlinecite{alba-2010,alba-2011,alba-2013}. More specifically, one can write
\begin{align}
\frac{Z(A_i,n,\beta)}{Z(A_{j},n,\beta)} = \frac{T_{j\rightarrow i}}{T_{i\rightarrow j}}, 
\end{align}
where $T_{i\rightarrow j}$  is the (Monte Carlo) transition probability from a spin 
configuration living on the booklet with subregion $A=A_i$ to a spin configuration living 
on the booklet with $A=A_j$. Notice that spins in region $A$ of different sheets are 
identified (cf. Eq.~\eqref{book-constraint}). Clearly, if $j<i$ one has $T_{i\rightarrow j}=1$.
When $j>i$ a naive method to determine $T_{i\rightarrow j}$ would be to simply count the 
fraction of times (during the Monte Carlo update) that a spin configuration living on 
the booklet with $A=A_i$ is also a valid spin configuration on the booklet with $A=A_j$.
In the following we provide a more efficient scheme to calculate $T_{i\rightarrow j}$.
To summarize, our approach for calculating the ratio of partition functions in 
Eq.~\eqref{ratio-trick} consists of four steps:
\begin{enumerate}
\item Do a full Monte Carlo sweep of the system on the booklet with $A=A_i$ 
to generate an importance sampled spin configuration. This can be done with any 
update scheme, such as standard Metropolis or cluster updates. 
\item For the spins living in the set difference $A_{i+1}\setminus A_i$ calculate 
\begin{align}
W_a = \prod_n\sum_{i_n} e^{-\beta E_{i_n}}.
\end{align}
Here (cf. Eq.~\eqref{sub-choice}) $i$ runs over all the $2^4$ configurations of the 
four spins in $A_{i+1}\setminus A_i$ for each of the $n$ sheets, and $E_{i_n}$ is 
the energy associated with the spin configuration $i_n$ on a single sheet. 
\item Calculate the quantity
\begin{align}
W_b = \sum_j e^{-\beta E_j}.
\end{align}
As in step $2$, $j$ runs over all the $2^4$ configurations of the four spins living 
in $A_{i+1}\setminus A_i$, but since the spins living on different sheets are identified 
there are only $2^4$ configurations in total and the energy $E_i$ is that of all $n$ 
sheets together.
\item Calculate the transition probability $T_{i\rightarrow i+1}$ as
\begin{align}
T_{i \rightarrow i+1} = \Bigl< \frac{W_b}{W_a} \Bigr>,
\end{align}
\end{enumerate}
where the angular brackets denote the Monte Carlo average. 
This method is much faster than simply counting configurations as it allows 
us to integrate over all possible configurations of the spins in $A_{i+1}\setminus A_i$, 
treating the remaining ones as a bath that is updated by the regular Monte Carlo update. 
The computational cost of this procedure grows exponentially with the number of spins 
in $A_{i+1}\setminus A_i$. 
In our Monte Carlo simulations we calculate $Z(A_i,n,\beta)/Z(A_{i+1},n,\beta)$ 
for every disorder realization and $S_n(A)$ using Eq.~\eqref{renyi-app}. 
Finally, we average over the disorder to obtain $[S_n(A)]$.


\end{document}